\documentclass[12pt,prd,tightenlines,nofootinbib,showpacs,showkeys]{revtex4}
\newcommand{\be}{\begin{equation}}
\newcommand{\ee}{\end{equation}}
\usepackage{bm}
\usepackage{graphics}
\usepackage{rotating}
\usepackage{epsfig}
\begin{document}
\title
{Spatial pattern formation in external noise: theory and simulation}
\author{\firstname{S.E.} \surname{Kurushina}}
\affiliation{Physics Department, Samara State Aerospace University named after S.P. Korolyov, Moskovskoye Shosse 34, 443086,
Samara, Russian Federation}
\author{\firstname{V.V.} \surname{Maximov}}
\affiliation{Physics Department, Samara State Aerospace University named after S.P. Korolyov, Moskovskoye Shosse 34, 443086,
Samara, Russian Federation}
\author{\firstname{Yu.M.} \surname{Romanovskii}}
\affiliation{Physics Department, Lomonosov Moscow State University, GSP-1, Leninskie Gory, 119991,
Moscow, Russian Federation}

\begin{abstract}
Spatial pattern formation in excitable fluctuating media was researched
analytically from the point of view of the order parameters concept. The
reaction-diffusion system in external noise is considered as a model of such
medium. Stochastic equations for the unstable mode amplitudes (order
parameters), dispersion equations for the unstable mode averaged amplitudes,
and the Fokker-Planck equation for the order parameters have been obtained. The
developed theory makes it possible to analyze different noise-induced
effects, including the variation of boundaries of ordering and disordering
phase transitions depending on the parameters of external noise
\end{abstract}

\pacs{05.40.Ca, 89.75.Kd, 02.50.Ey, 05.65.+b}

\keywords{Stochastic reaction-diffusion system, Spatial colored noise, Spatial pattern formation}

\maketitle

\section{Introduction}

Noise is present in real systems of any type. The influence of external
noise on nonlinear open distributed systems is very diverse and sometimes
leads to quite unexpected results. It is known, that noise-induced spatial
patterns \cite{GarHer1,ParBro2,GarSan3,ZaikSchim4,BucIba5,RiazDut6,SanzZhab7} and fronts \cite{Zhou8,SanSan9}, noise-induced resonant pattern and frequency-locking phenomena \cite{LiuJin10}, pure noise-induced phase transitions
\cite{KawSail11,BroPar12,BroPar13,GenSan14}, noise-induced phase separation \cite{IbaGar15}, noise-induced spatiotemporal
intermittency \cite{ZimTor16}, spatiotemporal stochastic resonance \cite{MarGam17,GamHan18,GosMar19},
noise-supported traveling and noise-sustained convective structures \cite{WanJun20,Deis21,SantCol22},
noise-induced synchronization \cite{SegShap23,NeiPei24,ElsSel25}, etc., may arise in such systems.

Theoretical study of spatio-temporal dynamics of nonlinear open distributed
systems is carried out using different methods \cite{LinGar26,GarSan27,HorLef28,Haken29,Gardiner30,Klyat31,ValSchi32,HutLong33,SanGar34}.

Stability of the homogeneous state with respect to small perturbations can
be analyzed in linear approximation. However, linear approximation
is not acceptable to describe the evolution of the system near the threshold
of self-organization, as the phenomenon self-organization itself is
essentially a nonlinear effect.

There is an approach to the study of noise-induced phenomena, based on the
well-known mean-field approximation \cite{ZaikSchim4,BroPar12,IbaGar15,LinGar26,GarSan27}. In this approximation
it is assumed, that the interaction between a certain spatial point and its
nearest neighbors occurs through the field, whose value corresponds to the
statistically average field at this point. Herewith, a suitable way is used
to carry out the discretization of the space of the initial distributed
system and the Fokker-Planck equation (FPE) for the multivariate probability
density function can be written for field values in the points received by a
regular lattice. The obtained FPE is integrated over the values of the field
at all points except the given one. This leads to FPE for the
one-dimensional probability density function values of the field at a given
point. In the latter equation the conditional average values of the field at
neighboring points are replaced by an average value of the field at a given
point. This approach can predict the existence of noise-induced
``disorder-order-disorder'' phase transitions.

Another analytical approach to the study of nonequilibrium phase transitions
with the spatial pattern formation is based on generalized Ginzburg-Landau
equations for the order parameters of the systems. This approach makes it
possible to describe the behavior of the system near the transition point.
It is based on the separation of eigenmodes of the system into damped
(stable) and undamped (unstable) modes (order parameters), and into
adiabatic elimination of stable modes. The method of generalized
Ginzburg-Landau equations for the systems, the right-hand side of which
contains additive white noise, is described in [29].

In papers \cite{GenSan14,SanGar34} spatio-temporal evolution of the nonequilibrium extended
systems is investigated by means of dynamic renormalization groups. In \cite{GenSan14}
it is shown that under certain conditions a new genuine nonequilibrium
universality class arises due to the presence of multiplicative noise.

Moreover, there are approaches based on analysis of correlation functions
of the dynamic variables of the system or structure functions \cite{GarSan3,BroPar12}, on the
study of higher-order moments \cite{RiazDut6,ValSchi32}, and others \cite{Gardiner30,Klyat31,HutLong33,Roman40}.

The aim of this paper is to develop a theory, which would allow us from the
unified point of view of the concept of order parameters, to carry out a
consistent and detailed study of spatial pattern formation spontaneously
arising in open nonlinear distributed systems with external noise both in
the vicinity of the transition point and away from it.

\section{Stochastic equations for the order parameters}

Systems of the reaction-diffusion type

\begin{equation}
\label{eq1}
{\frac{{\partial x_{k}}} {{\partial t}}} = P_{k} (x_{1} ,x_{2} ,x_{3}
,...,  {\mbox{\boldmath$\eta$}},{\rm {\bf r}}, t) + D_{k} \nabla ^{2}x_{k}, k = 1,2,3\ldots
\end{equation}

\noindent
are one of the basic models of a nonequilibrium excitable medium.
In equation (\ref{eq1}) $x_{k}$ are the medium state functions (dynamic variables),
$P_{k} (x_{1} ,x_{2} ,x_{3} ,...,{\mbox{\boldmath$\eta$}},{\rm {\bf r}},t)$ are
nonlinear functions, that define the interaction and the evolution of the
component $x_{k}$ in space and in time, $D_{k}$ are diffusion coefficients of
components, {\boldmath{$\eta$}} = (\textit{$\chi $}$_{{\rm 1}}$,\ldots ,\textit{$\chi $}$_{n}$,\textit{$\eta $}$_{{\rm1}}$,\ldots ,\textit{$\eta $}$_{s})$ is the vector, whose components are the control
parameters describing the effect of the external environment on the system.
Further, without loss of generality, we consider a two-component system of
type (\ref{eq1}). Nevertheless, the proposed research method is easily extended to
multicomponent systems.

In external noise the spatio-temporal dynamics of nonequilibrium systems (\ref{eq1})
for k = 2 can be described by the following system of equations:

\begin{equation}
\label{eq2}
\begin{array}{c}
{\frac{{\partial x_{1}}} {{\partial t}}} =
 P_{1} (x_{1} ,x_{2} ,\chi _{10},...,\chi _{m0} ,...,\chi _{n} ) +
{\sum\limits_{j = 1}^{m} {f_{1j} ({\rm {\bf r}},t)P_{1j} (x_{1} ,x_{2} ,\chi _{m + 1} ,...,\chi _{{n}'} )}}  +
D_{1} \nabla ^{2}x_{1}, \\
{\frac{{\partial x_{2}}} {{\partial t}}} =
P_{2} (x_{1} ,x_{2} ,\eta _{10},...,\eta _{l0} ,...,\eta _{s} ) +
{\sum\limits_{j = 1}^{l} {f_{2j} ({\rm {\bf r}},t)P_{2j} (x_{1} ,x_{2} ,\eta _{l + 1} ,...,\eta _{{s}'} )}}  +
D_{2} \nabla ^{2}x_{2},
\end{array}
\end{equation}

\noindent
where $m$ and $l$ are the number of fluctuating parameters in the first and second
equations, respectively, \textit{$\chi $}$_{j{\rm 0}}$, \textit{$\eta $}$_{j{\rm 0}}$ are spatio-temporal
average parameters, $f_{ij}$(\textbf{r},$t)$ ($i$=1,2) are random fields describing
the noise of the appropriate parameters with respect to their mean values
with ${\left\langle {f_{ij} ({\rm {\bf r}},t)} \right\rangle}  = 0$.

We define the statistical properties of random fields $f_{ij}$(\textbf{r},$t)$
according to the properties of the environment. Fluctuations in the
environment represent the summarized effect of many weakly coupled factors.
It follows from the central limit theorem, that fluctuations of the external
source have a Gaussian distribution. The ergodic Markovian and Gaussian
properties of fluctuating environment limits the choice of random fields for
modeling the fluctuations of the environment by a stationary homogeneous
isotropic Gaussian field with the exponential time-correlation function
\cite{HorLef28}:

\begin{equation}
\label{eq3}
K{\left[ {f_{ij} ({\rm {\bf r}},t),f_{{i}'{j}'} ({\rm {\bf {r}'}},{t}')}
\right]} = \Phi _{i} ({\left| {{\rm {\bf r}} - {\rm {\bf {r}'}}}
\right|})\exp ( - k_{ti} {\left| {t - {t}'} \right|})\delta _{i{i}'} \delta
_{j{j}'} ,
\end{equation}

\noindent
where $\Phi _{i} ({\left| {{\rm {\bf r}} - {\rm {\bf {r}'}}} \right|})$
define the spatial dependence of correlation functions of the random fields.
The cross-correlation of the fields $f_{ij} ({\rm {\bf r}},t)$ and
$f_{{i}'{j}'} ({\rm {\bf r}},t)$ is absent. The correlation time $r_{t} =
k_{t}^{ - 1} $ is the characteristic time scale of external noise. Hereafter,
we use the notation $K[F_{1} ,F_{2} ]$, that is defined by the equality $K[F_{1}
,F_{2} ] =  \langle F_{1} F_{2} \rangle - \langle F_{1} \rangle \langle F_{2} \rangle$ for the correlation function.

Let us assume for simplicity, that $m=l$ = 1, and introduce the dimensionless
variables $\tau = \chi _{10} t$ and ${\rm {\bf {r}'}} = {\rm {\bf r}}\sqrt
{{{\chi _{10}}  \mathord{\left/ {\vphantom {{\chi _{10}}  {D_{1}}} } \right.
\kern-\nulldelimiterspace} {D_{1}}} } $, where \textit{$\chi $}$_{{\rm 1}{\rm 0}}$ is a
parameter, that has the dimension of inverse time. Now the system (\ref{eq2}) can be
rewritten as

\begin{equation}
\label{eq4}
\begin{array}{c}
{\frac{{\partial x_{1}}} {{\partial \tau}} } = {P}'_{1} (x_{1} ,x_{2} ,\chi
_{10} ,...,\chi _{n} ) + f_{11} ({\rm {\bf {r}'}},\tau ){P}'_{11} (x_{1}
,x_{2} ,\chi _{2} ,...,\chi _{{n}'} ) + {\nabla} '^{2}x_{1} , \\
{\frac{{\partial x_{2}}} {{\partial \tau}} } = {P}'_{2} (x_{1} ,x_{2} ,\eta
_{10} ,...,\eta _{s} ) + f_{21} ({\rm {\bf {r}'}},\tau ){P}'_{21} (x_{1}
,x_{2} ,\eta _{2} ,...,\eta _{{s}'} ) + D{\nabla} '^{2}x_{2}.
\end{array}
\end{equation}

\noindent Functions with primes are different from the corresponding functions without
primes by a multiplier $\chi _{10}^{ - 1} $. $D = D_{2} / D_{1} $. Hereafter,
the primes are dropped for simplicity.

Suppose, that in the deterministic case the values of parameter \textit{$\chi $}$_{{\rm
1}}$,\ldots ,\textit{$\chi $}$_{n}$,\textit{$\eta $}$_{{\rm 1}}$,\ldots ,\textit{$\eta $}$_{s}$ are such, that there are stable stationary states $x_{{\rm 1}{\rm 0}}$ and $x_{{\rm 2}{\rm 0}}$ defined
by the equations $P_{1} (x_{1} ,x_{2} ,\chi _{1} ,...,\chi _{n} ) = 0$ and
$P_{2} (x_{1} ,x_{2} ,\eta _{1} ,...,\eta _{s} ) = 0$.

We write the equations (\ref{eq4}) in the operator form. Simultaneously, we select
from its right-hand side linear $K(\nabla ^{2}){\rm {\bf q}}$, nonlinear
${\rm {\bf g}}({\rm {\bf q}})$, and random\textbf{ F}(\textbf{r}\textit{,$\tau $}) components
\cite{Haken29}

\begin{equation}
\label{eq5}
{\frac{{\partial {\rm {\bf q}}}}{{\partial \tau}} } - K(\nabla ^{2}){\rm
{\bf q}} = {\rm {\bf g}}({\rm {\bf q}}) + {\rm {\bf F}}({\rm {\bf r}},\tau).
\end{equation}

Vector \textbf{q} describes the deviation of the dynamic variables from
their equilibrium values: ${\rm {\bf q}} = (x_{1} - x_{10} ,x_{2} - x_{20}
)$. A linear operator $K(\nabla ^{2})$ takes the form

\begin{equation}
\label{eq6}
 K(\nabla ^{2}) = \left( {{\begin{array}{cc}
 {a_{11} + \nabla ^{2}} \hfill & {a_{12}}  \hfill \\
 {a_{21}}  \hfill & {a_{22} + D\nabla ^{2}} \hfill \\
\end{array}}}  \right),
\quad
a_{ij} = {\left. {{\frac{{\partial P_{i}}} {{\partial x_{j}}} }}
\right|}_{x_{10} ,x_{20}}  , \quad  i,j = 1,2.
\end{equation}

Vector \textbf{g}(\textbf{q}) contains quadratic and cubic nonlinearities
obtained by series expansion of the right-hand deterministic side of the
equation (\ref{eq4}). Its components are defined as follows

\begin{equation}
\label{eq7}
g_{i} ({\rm {\bf q}}) = {\sum\limits_{\mu ,\nu = 1}^{2} {g_{i,\mu \nu
}^{(2)} q_{\mu}  q_{\nu}} }   + {\sum\limits_{\mu ,\nu ,\kappa = 1}^{2}
{g_{i,\mu \nu \kappa} ^{(3)} q_{\mu}  q_{\nu}  q_{\kappa}} }  ,
\end{equation}

\noindent
where $g_{i,\mu \nu} ^{(2)} = {\frac{{1}}{{2!}}}{\left.
{{\frac{{\partial ^{2}P_{i}}} {{\partial x_{\mu}  \partial x_{\nu}} } }}
\right|}_{x_{10} ,x_{20}}  $, $g_{i,\mu \nu \kappa} ^{(3)} =
{\frac{{1}}{{3!}}}{\left. {{\frac{{\partial ^{3}P_{i}}} {{\partial x_{\mu}
\partial x_{\nu}  \partial x_{\kappa}} } }} \right|}_{x_{10} ,x_{20}} $.

Vector \textbf{F} contains the random fields:
\[
{\rm {\bf F}} = \left( {{\begin{array}{*{20}c}
 {f_{11} ({\rm {\bf r}},\tau )P_{11} (x_{1} ,x_{2} ,\chi _{2} ,...,\chi
_{{n}'} )} \hfill \\
 {f_{21} ({\rm {\bf r}},\tau )P_{21} (x_{1} ,x_{2} ,\eta _{2} ,...,\eta
_{{s}'} )} \hfill \\
\end{array}}}  \right).
\]

To research the stability of the stationary state of a deterministic system,
we assume that the vector \textbf{q} has the form ${\rm {\bf q}} = {\rm {\bf
q}}_{0} \exp (\lambda \tau + i{\rm {\bf k}}{\rm {\bf r}})$. The respective
characteristic equation $\lambda ^{2} - \alpha \lambda + \beta = 0$ has
solutions

\begin{equation}
\label{eq8}
\lambda _{1,2} ({\rm {\bf k}}) = {\frac{{\alpha (k)\pm \sqrt {\alpha ^{2}(k)
- 4\beta (k)}}} {{2}}},
\end{equation}

\noindent
where $\alpha (k) = Tr(a_{ij} ) - (1 + D)k^{2}$, $\beta (k) = Det(a_{ij} ) -
(Da_{11} + a_{22} )k^{2} + Dk^{4}$.

Conditions $\alpha (k) < 0$ and $\beta (k) \le 0$ define aperiodic
instability, herewith one positive real root of eq. (\ref{eq8}) appears: $Re(\lambda
_{1} ) \ge 0$, $Im(\lambda _{1,2} ) = 0$. Further we shall consider only
this case.

Suppose, that one of the parameters \textit{$\chi $}$_{{\rm 1}}$,\ldots ,\textit{$\chi $}$_{n}$,\textit{$\eta $}$_{{\rm
1}}$,\ldots ,\textit{$\eta $}$_{s}$, for example $\chi _{{\rm 2}}$, is bifurcation, i.e.,
at some critical value of this parameter \textit{$\chi $}$_{{\rm 2 c}}$ there exists a
critical wave number k$_{{\rm c}}$, when the conditions $Re(\lambda _{1}
(k_{c} )) = 0$, ${\left. {{\frac{{d(Re\lambda _{1} (k))}}{{dk}}}}
\right|}_{k = kc} = 0$ are fulfilled.

Represent the vector \textbf{q} in the form of superposition

\begin{equation}
\label{eq9}
{\rm {\bf q}}({\rm {\bf r}},\tau ) = {\sum\limits_{{\rm {\bf {k}'}},j} {{\rm
{\bf O}}^{(j)}({\rm {\bf {k}'}})\xi _{{\rm {\bf {k}'}}}^{(j)} (\tau
)e^{i{\rm {\bf {k}'}}{\rm {\bf r}}}}} ,
\end{equation}

\noindent
where ${\rm {\bf O}}^{(j)}({\rm {\bf k}})$ are eigenvectors of operator
$K(\nabla ^{2})$, $\xi _{{\rm {\bf k}}}^{(j)} (\tau )$ are unknown
amplitudes, $\exp (i{\rm {\bf k}}{\rm {\bf r}})$ are eigenfunctions of
operator $\nabla ^{2}$. Here it is assumed, that the vector
\textbf{q}(\textbf{r},\textit{$\tau $}) is a superposition of plane waves, but depending on
the symmetry of the problem Bessel functions or spherical wave functions are
to be chosen as the eigenfunctions.

The vector \textbf{F} contains nonlinear functions $P_{i{\rm 1}}$. They need
to be expanded in a Taylor series in powers of the components of vector
\textbf{q}. We restrict our consideration to quadratic terms in the
expansion of $P_{i{\rm 1}}$. It is easily shown, that the terms of this order
are necessary to obtain the dispersion equation for the averaged amplitudes
of the unstable modes with an accuracy to terms quadratic in the
fluctuation intensity. As a result, the components of vector F containing
random fields take the form:

\[
F_{i} = f_{i1} ({\rm {\bf r}},\tau )\left( {p_{i}^{(0)} + {\sum\limits_{\mu
= 1}^{2} {p_{i,\mu} ^{(1)} q_{\mu}  + {\sum\limits_{\mu ,\nu = 1}^{2}
{p_{i,\mu \nu} ^{(2)} q_{\mu}  q_{\nu}} } } }}  \right),
\]

\noindent
where $p_{i}^{(0)} = P_{i1} (x_{10} ,x_{20} )$, $p_{i,\mu} ^{(1)} = {\left.
{{\frac{{\partial P_{i1}}} {{\partial x_{\mu}} } }} \right|}_{x_{10} ,x_{20}
} $, $p_{i,\mu \nu} ^{(2)} = {\frac{{1}}{{2!}}}{\left. {{\frac{{\partial
^{2}P_{i1}}} {{\partial x_{\mu}  \partial x_{\nu}} } }} \right|}_{x_{10}
,x_{20}}  $.

Unstable modes lie in a narrow band of wavenumbers defining the instability
region of the system. This makes it possible to construct wave packets by
summing over the wave vectors, which lie in a small interval. Thus, carrying
modes with discrete values of wave vectors and slowly varying amplitudes
$\xi _{{\rm {\bf k}}}^{(j)} (\tau )$ are chosen \cite{Haken29}.

To obtain the equations for the mode amplitudes $\xi _{{\rm {\bf k}}}^{(j)}
(\tau )$ we substitute eq. (\ref{eq9}) in eq. (\ref{eq5}), multiply the equation derived on
the left by $\exp ( - i{\rm {\bf k}}{\rm {\bf r}}){\rm {\bf O}}^{ *
({j}')}({\rm {\bf k}})$ and integrate over the region, which is much greater
than the oscillation period $\exp (i{\rm {\bf k}}{\rm {\bf r}})$, but in
which $\xi _{{\rm {\bf k}}}^{(j)} (\tau )$ varies very little. Here ${\rm
{\bf O}}^{ * ({j}')}({\rm {\bf k}})$ are eigenvectors of the operator
conjugate to $K(\nabla ^{2})$: ${\rm {\bf O}}^{(j)}{\rm {\bf O}}^{ * ({j}')}
= \delta _{j{j}'} $.

After transformations, the system of equations for the amplitudes of the
modes takes the form

\begin{equation}
\label{eq10}
{\frac{{d\xi _{{\rm {\bf k}}}^{(j)}}} {{d\tau}} } - \lambda _{j} ({\rm {\bf
k}})\xi _{{\rm {\bf k}}}^{(j)} = {\sum\limits_{{j}'{j}'',{\rm {\bf
{k}'}}{\rm {\bf {k}''}}} {\sigma _{{j}'{j}''}^{(j)} ({\rm {\bf k}},{\rm {\bf
{k}'}},{\rm {\bf {k}''}})}} \xi _{{\rm {\bf {k}'}}}^{({j}')} \xi _{{\rm {\bf
{k}''}}}^{({j}'')} \delta ({\rm {\bf {k}'}} + {\rm {\bf {k}''}} - {\rm {\bf
k}}) +
\end{equation}
\begin{displaymath}
 + {\sum\limits_{{j}'{j}''{j}''',{\rm {\bf {k}'}}{\rm {\bf {k}''}}{\rm {\bf
{k}'''}}} {\sigma _{{j}'{j}''{j}'''}^{(j)} ({\rm {\bf k}},{\rm {\bf
{k}'}},{\rm {\bf {k}''}},{\rm {\bf {k}'''}})}} \xi _{{\rm {\bf
{k}'}}}^{({j}')} \xi _{{\rm {\bf {k}''}}}^{({j}'')} \xi _{{\rm {\bf
{k}'''}}}^{({j}''')} \delta ({\rm {\bf {k}'}} + {\rm {\bf {k}''}} + {\rm
{\bf {k}'''}} - {\rm {\bf k}}) + {\sum\limits_{\varphi = 1}^{2} {O_{\varphi
}^{ * (j)} ({\rm {\bf k}})p_{\varphi} ^{(0)}}}  z_{\varphi ,{\rm {\bf k}}}
(\tau ) +
\end{displaymath}
\begin{displaymath}
+ {\sum\limits_{\varphi = 1}^{2} {{\sum\limits_{{j}',{\rm {\bf {k}'}}}
{\varepsilon _{\varphi ,{j}'}^{(j)} ({\rm {\bf k}},{\rm {\bf {k}'}})\xi
_{{\rm {\bf {k}'}}}^{({j}')} z_{\varphi ,{\rm {\bf k}} - {\rm {\bf {k}'}}}
(\tau )}}} }  + {\sum\limits_{\varphi = 1}^{2} {{\sum\limits_{{j}'{j}'',{\rm
{\bf {k}'}}{\rm {\bf {k}''}}} {\varepsilon _{\varphi ,{j}'{j}''}^{(j)} ({\rm
{\bf k}},{\rm {\bf {k}'}},{\rm {\bf {k}''}})\xi _{{\rm {\bf {k}'}}}^{({j}')}
\xi _{{\rm {\bf {k}''}}}^{({j}'')} z_{\varphi ,{\rm {\bf k}} - {\rm {\bf
{k}'}} - {\rm {\bf {k}''}}} (\tau )}}} } .
\end{displaymath}

Functions $\sigma _{{j}'{j}''}^{(j)} ({\rm {\bf k}},{\rm {\bf {k}'}},{\rm
{\bf {k}''}})$, $\sigma _{{j}'{j}''{j}'''}^{(j)} ({\rm {\bf k}},{\rm {\bf
{k}'}},{\rm {\bf {k}''}},{\rm {\bf {k}'''}})$, $\varepsilon _{\varphi
,{j}'}^{(j)} ({\rm {\bf k}},{\rm {\bf {k}'}})$, $\varepsilon _{\varphi
,{j}'{j}''}^{(j)} ({\rm {\bf k}},{\rm {\bf {k}'}},{\rm {\bf {k}''}})$
introduced in equations (\ref{eq10}) are presented in the Appendix A.

Random processes $z_{\varphi,{\rm {\bf k}}}(\tau )=\int {f_{\varphi 1}({\rm
{\bf r}},\tau )e^{ - i{\rm {\bf k}}{\rm {\bf r}}}d{\rm {\bf r}}} $ are
components of the random vector field \textbf{z}(\textit{$\tau $}) with zero mean, \textit{$\varphi$} and
\textbf{k} are index arguments of this field.

Assuming, that the correlation time of random fields is considerably smaller
than all the characteristic times of the deterministic problem (\ref{eq4}),
correlation functions for the components of field \textbf{z}(\textit{$\tau $}) will have the
form: $K[z_{j,{\rm {\bf k}}} (t),z_{l,{\rm {\bf {k}'}}} (\tau )] = g_{jl}
({\left| {{\rm {\bf k}}} \right|})\delta ({\rm {\bf k}} - {\rm {\bf
{k}'}})\delta (t - \tau )\delta _{jl} $. Further we assume for definiteness

\begin{equation}
\label{eq11}
\Phi _{j} ({\left| {{\rm {\bf r}} - {\rm {\bf {r}'}}} \right|}) = \theta
_{j} \exp ( - k_{fj} {\left| {{\rm {\bf r}} - {\rm {\bf {r}'}}}
\right|}).
\end{equation}

\noindent
Here \textit{$\theta $}$_{j}$ are noise intensity, $k_{fj}$ are magnitudes inverse to
correlation length. For two-dimensional media $g_{jj} = \theta _{j} k_{fj} /
[2\pi ^{2}(k^{2} + k_{fj}^{2} )^{ - 3 / 2}]$.

The system (\ref{eq10}) contains both stable and unstable modes. In the vicinity of
the bifurcation point the relaxation times of unstable modes are
considerably greater then the relaxation times of stable modes, therefore
the latter adiabatically follow the former ones. This makes it possible to
exclude the stable modes from the equations (\ref{eq10}). To perform the procedure
of adiabatic elimination \cite{Haken29,Gardiner30} of stable modes, we rewrite the system of
equations (\ref{eq10}) dividing it into two subsystems of equations for the
unstable modes (denote them by an additional index (u)) and for the stable
modes (s).

Since the unstable modes can grow to infinity if we neglect the nonlinear
terms, we write equations for them with an accuracy of cubic terms, which
provide nonlinear stabilization of the instability (if the cubic terms are
not enough, it is necessary to take into account the fifth orders).

We assume that the amplitudes of stable modes are significantly smaller than
the amplitudes of unstable modes ${\left| {\xi _{s}}  \right|} < < {\left|
{\xi _{u}}  \right|}$ and their variations occur self-consistently: $\xi
_{s} \sim \xi _{u}^{2} $. In addition, $ z_{\varphi,{\rm {\bf k}}} \sim \xi_{s}$.

In the equations for the stable mode amplitudes we keep only the terms
necessary to obtain the equations for the unstable mode amplitudes with an
accuracy to third-order terms. Then from eq. (\ref{eq10}) for the stable mode
amplitudes we obtain the following equation:

\begin{equation}
\label{eq12}
{\frac{{d\xi _{{\rm {\bf k}}s}^{(j)}}} {{d\tau}} } - \lambda _{j} ({\rm {\bf
k}}_{s} )\xi _{{\rm {\bf k}}s}^{(j)} = {\sum\limits_{{\rm {\bf {k}'}}u{\rm
{\bf {k}''}}u} {\sigma _{11}^{(j)} ({\rm {\bf k}}_{s} ,{\rm {\bf {k}'}}_{u}
,{\rm {\bf {k}''}}_{u} )}} \xi _{{\rm {\bf {k}'}}u}^{(1)} \xi _{{\rm {\bf
{k}''}}u}^{(1)} \delta ({\rm {\bf {k}'}}_{u} + {\rm {\bf {k}''}}_{u} - {\rm
{\bf k}}_{s} ) +
\end{equation}
\begin{displaymath}
 + {\sum\limits_{\varphi = 1}^{2} {O_{\varphi} ^{ * (j)} ({\rm {\bf k}}_{s}
)p_{\varphi} ^{(0)}}}  z_{\varphi ,{\rm {\bf k}}s} (\tau ) +
{\sum\limits_{\varphi = 1}^{2} {{\sum\limits_{{\rm {\bf {k}'}}u}
{\varepsilon _{\varphi ,1}^{(j)} ({\rm {\bf k}}_{s} ,{\rm {\bf {k}'}}_{u}
)\xi _{{\rm {\bf {k}'}}u}^{(1)} z_{\varphi ,{\rm {\bf k}}s - {\rm {\bf
{k}'}}u} (\tau ) +}} } }
\end{displaymath}
\begin{displaymath}
 + {\sum\limits_{\varphi = 1}^{2} {{\sum\limits_{{\rm {\bf {k}'}}u{\rm {\bf
{k}''}}u} {\varepsilon _{\varphi ,11}^{(j)} ({\rm {\bf k}}_{s} ,{\rm {\bf
{k}'}}_{u} ,{\rm {\bf {k}''}}_{u} )\xi _{{\rm {\bf {k}'}}u}^{(1)} \xi _{{\rm
{\bf {k}''}}u}^{(1)} z_{\varphi ,{\rm {\bf k}}s - {\rm {\bf {k}'}}u - {\rm
{\bf {k}''}}u} (\tau )}}} } .
\end{displaymath}

The equations for the unstable mode amplitudes have the form:

\begin{equation}
\label{eq13}
{\frac{{d\xi _{{\rm {\bf k}}u}^{(1)}}} {{d\tau}} } - \lambda _{1} ({\rm {\bf
k}}_{u} )\xi _{{\rm {\bf k}}u}^{(1)} = {\sum\limits_{{\rm {\bf {k}'}}u{\rm
{\bf {k}''}}u} {\sigma _{11}^{(1)} ({\rm {\bf k}}_{u} ,{\rm {\bf {k}'}}_{u}
,{\rm {\bf {k}''}}_{u} )}} \xi _{{\rm {\bf {k}'}}u}^{(1)} \xi _{{\rm {\bf
{k}''}}u}^{(1)} \delta ({\rm {\bf {k}'}}_{u} + {\rm {\bf {k}''}}_{u} - {\rm
{\bf k}}_{u} ) +
\end{equation}
\begin{displaymath}
 + {\sum\limits_{{\rm {\bf {k}'}}u{\rm {\bf {k}''}}u{\rm {\bf {k}'''}}u}
{\sigma _{111}^{(1)} ({\rm {\bf k}}_{u} ,{\rm {\bf {k}'}}_{u} ,{\rm {\bf
{k}''}}_{u} ,{\rm {\bf {k}'''}}_{u} )}} \xi _{{\rm {\bf {k}'}}u}^{(1)} \xi
_{{\rm {\bf {k}''}}u}^{(1)} \xi _{{\rm {\bf {k}'''}}u}^{(1)} \delta ({\rm
{\bf {k}'}}_{u} + {\rm {\bf {k}''}}_{u} + {\rm {\bf {k}'''}}_{u} - {\rm {\bf
k}}_{u} ) +
\end{displaymath}
\begin{displaymath}
 + {\sum\limits_{\varphi = 1}^{2} {O_{\varphi} ^{ * (1)} ({\rm {\bf k}}_{u}
)p_{\varphi} ^{(0)}}}  z_{\varphi ,{\rm {\bf k}}u} (\tau ) +
{\sum\limits_{\varphi = 1}^{2} {{\sum\limits_{{\rm {\bf {k}'}}u}
{\varepsilon _{\varphi ,1}^{(1)} ({\rm {\bf k}}_{u} ,{\rm {\bf {k}'}}_{u}
)\xi _{{\rm {\bf {k}'}}u}^{(1)} z_{\varphi ,{\rm {\bf k}}u - {\rm {\bf
{k}'}}u} (\tau )}}} }  +
\end{displaymath}
\begin{displaymath}
+ {\sum\limits_{\varphi = 1}^{2} {{\sum\limits_{{\rm {\bf {k}'}}u{\rm {\bf
{k}''}}u} {\varepsilon _{\varphi ,11}^{(1)} ({\rm {\bf k}}_{u} ,{\rm {\bf
{k}'}}_{u} ,{\rm {\bf {k}''}}_{u} )}} \xi _{{\rm {\bf {k}'}}u}^{(1)} \xi
_{{\rm {\bf {k}''}}u}^{(1)}}}  z_{\varphi ,{\rm {\bf k}}u - {\rm {\bf
{k}'}}u - {\rm {\bf {k}''}}u} (\tau )+
\end{displaymath}
\begin{displaymath}
+ {\sum\limits_{\psi ,\varphi = 1}^{2}
{{\sum\limits_{{\rm {\bf k}}s} {\varepsilon _{\varphi ,\psi} ^{(1)} ({\rm
{\bf k}}_{u} ,{\rm {\bf k}}_{s} )\xi _{{\rm {\bf k}}s}^{(\psi )} z_{\varphi
,{\rm {\bf k}}u - {\rm {\bf k}}s} (\tau )}}} }  +
\end{displaymath}
\begin{displaymath}
+ {\sum\limits_{\psi = 1}^{2} {{\sum\limits_{{\rm {\bf {k}'}}u{\rm {\bf k}}s}
{{\left[ {\sigma _{1\psi} ^{(1)} ({\rm {\bf k}}_{u} ,{\rm {\bf {k}'}}_{u}
,{\rm {\bf k}}_{s} ) + \sigma _{\psi 1}^{(1)} ({\rm {\bf k}}_{u} ,{\rm {\bf
k}}_{s} ,{\rm {\bf {k}'}}_{u} )} \right]}}} \xi _{{\rm {\bf {k}'}}u}^{(1)}
\xi _{{\rm {\bf k}}s}^{(\psi )} \delta ({\rm {\bf {k}'}}_{u} + {\rm {\bf
k}}_{s} - {\rm {\bf k}}_{u} )}}  +
\end{displaymath}
\begin{displaymath}
 + {\sum\limits_{\psi ,\varphi = 1}^{2} {{\sum\limits_{{\rm {\bf {k}'}}u{\rm
{\bf k}}s} {{\left[ { \varepsilon _{\varphi ,1\psi} ^{(1)} ({\rm {\bf
k}}_{u} ,{\rm {\bf {k}'}}_{u} ,{\rm {\bf k}}_{s} ) + \varepsilon _{\varphi
,\psi 1}^{(1)} ({\rm {\bf k}}_{u} ,{\rm {\bf k}}_{s} ,{\rm {\bf {k}'}}_{u}
)} \right]}}} \xi _{{\rm {\bf {k}'}}u}^{(1)} \xi _{{\rm {\bf k}}s}^{(\psi )}
z_{\varphi ,{\rm {\bf k}}u - {\rm {\bf {k}'}}u - {\rm {\bf k}}s} (\tau )}
}.
\end{displaymath}

Neglecting the time derivative $d\xi _{{\rm {\bf k}}s}^{(j)} / d\tau $ in
equations (\ref{eq12}) \cite{Haken29}, expressing the amplitudes $\xi _{{\rm {\bf k}}s}^{(j)} $
from them and substituting the latter in eq. (\ref{eq13}), we obtain a system of
equations for the unstable mode amplitudes $\xi _{{\rm {\bf k}}u}^{(1)} $:

\begin{equation}
\label{eq14}
{\frac{{d\xi _{{\rm {\bf k}}u}^{(1)}}} {{d\tau}} } = F_{{\rm {\bf k}}u}
(\tau ),
\end{equation}

\[
F_{{\rm {\bf k}}u} (\tau ) = \lambda _{1} ({\rm {\bf k}}_{u} )\xi _{{\rm
{\bf k}}u}^{(1)} + {\sum\limits_{{\rm {\bf {k}'}}u{\rm {\bf {k}''}}u}
{\sigma _{11}^{(1)} ({\rm {\bf k}}_{u} ,{\rm {\bf {k}'}}_{u} ,{\rm {\bf
{k}''}}_{u} )}} \xi _{{\rm {\bf {k}'}}u}^{(1)} \xi _{{\rm {\bf
{k}''}}u}^{(1)} \delta ({\rm {\bf {k}'}}_{u} + {\rm {\bf {k}''}}_{u} - {\rm
{\bf k}}_{u} ) +
\]
\[
 + {\sum\limits_{{\rm {\bf {k}'}}u{\rm {\bf {k}''}}u{\rm {\bf {k}'''}}u}
{\omega ({\rm {\bf k}}_{u} ,{\rm {\bf {k}'}}_{u} ,{\rm {\bf {k}''}}_{u}
,{\rm {\bf {k}'''}}_{u} ,{\rm {\bf k}}_{u} - {\rm {\bf {k}'}}_{u} )}} \xi
_{{\rm {\bf {k}'}}u}^{(1)} \xi _{{\rm {\bf {k}''}}u}^{(1)} \xi _{{\rm {\bf
{k}'''}}u}^{(1)} \delta ({\rm {\bf {k}'}}_{u} + {\rm {\bf {k}''}}_{u} + {\rm
{\bf {k}''}}_{u} - {\rm {\bf k}}_{u} ) +
\]
\[
 + {\sum\limits_{\varphi = 1}^{2} {O_{\varphi} ^{ * (1)} ({\rm {\bf k}}_{u}
)p_{\varphi} ^{(0)}}}  z_{\varphi ,{\rm {\bf k}}u} (\tau ) -
{\sum\limits_{\psi ,\varphi ,{\varphi} ' = 1}^{2} {{\sum\limits_{{\rm {\bf
k}}s} {\zeta _{\varphi ,\psi ,{\varphi} '} ({\rm {\bf k}}_{u} ,{\rm {\bf
k}}_{s} )z_{\varphi ,{\rm {\bf k}}u - {\rm {\bf k}}s} (\tau )}}}
}z_{{\varphi} ',{\rm {\bf k}}s} (\tau ) +
\]
\[
 + {\sum\limits_{\varphi = 1}^{2} {{\sum\limits_{{\rm {\bf {k}'}}u} {\eta
_{\varphi}  ({\rm {\bf k}}_{u} ,{\rm {\bf {k}'}}_{u} )\xi _{{\vec
{k}}'u}^{(1)} z_{\varphi ,{\rm {\bf k}}u - {\rm {\bf {k}'}}u} (\tau )}}} }
 + {\sum\limits_{\varphi = 1}^{2} {{\sum\limits_{{\rm {\bf {k}'}}u{\rm {\bf
{k}''}}u} {\nu _{\varphi}  ({\rm {\bf k}}_{u} ,{\rm {\bf {k}'}}_{u} ,{\rm
{\bf {k}''}}_{u} )}} \xi _{{\rm {\bf {k}'}}u}^{(1)} \xi _{{\rm {\bf
{k}''}}u}^{(1)}}}  z_{\varphi ,{\rm {\bf k}}u - {\rm {\bf {k}'}}u - {\rm
{\bf {k}''}}u} (\tau ) -
\]
\[
 - {\sum\limits_{\psi ,\varphi ,{\varphi} ' = 1}^{2} {{\sum\limits_{{\rm
{\bf k}}s,{\rm {\bf {k}'}}u} {A_{\varphi ,\psi ,{\varphi} '} ({\rm {\bf
k}}_{u} ,{\rm {\bf k}}_{s} ,{\rm {\bf {k}'}}_{u} )\xi _{{\rm {\bf
{k}'}}u}^{(1)} z_{{\varphi} ',{\rm {\bf k}}s - {\rm {\bf {k}'}}u} (\tau
)z_{\varphi ,{\rm {\bf k}}u - {\rm {\bf k}}s} (\tau )}}} }  -
\]
\[
 - {\sum\limits_{\psi ,\varphi ,{\varphi} ' = 1}^{2} {{\sum\limits_{{\rm
{\bf {k}'}}u{\rm {\bf k}}s} {B_{\varphi ,\psi ,{\varphi} '} ({\rm {\bf
k}}_{u} ,{\rm {\bf k}}_{s} ,{\rm {\bf {k}'}}_{u} )}} \xi _{{\rm {\bf
{k}'}}u}^{(1)} z_{\varphi ,{\rm {\bf k}}u - {\rm {\bf {k}'}}u - {\rm {\bf
k}}s} (\tau )z_{{\varphi} ',{\rm {\bf k}}s} (\tau )}}  -
\]
\[
 - {\sum\limits_{\psi ,\varphi ,{\varphi} ' = 1}^{2} {{\sum\limits_{{\rm
{\bf {k}'}}u{\rm {\bf {k}''}}u{\rm {\bf k}}s} {C_{\varphi ,\psi ,{\varphi
}'} ({\rm {\bf k}}_{u} ,{\rm {\bf k}}_{s} ,{\rm {\bf {k}'}}_{u} ,{\rm {\bf
{k}''}}_{u} )}} \xi _{{\rm {\bf {k}'}}u}^{(1)} \xi _{{\rm {\bf
{k}''}}u}^{(1)} z_{{\varphi} ',{\rm {\bf k}}s - {\rm {\bf {k}'}}u - {\rm
{\bf {k}''}}u} (\tau )z_{\varphi ,{\rm {\bf k}}u - {\rm {\bf k}}s} (\tau )}
} -
\]
\[
 - {\sum\limits_{\psi ,{\varphi} ',\varphi = 1}^{2} {{\sum\limits_{{\rm {\bf
{k}'}}u{\rm {\bf {k}''}}u{\rm {\bf k}}s} {D_{\varphi ,\psi ,{\varphi} '}
({\rm {\bf k}}_{u} ,{\rm {\bf {k}'}}_{u} ,{\rm {\bf k}}_{s} ,{\rm {\bf
{k}''}}_{u} )}} \xi _{{\rm {\bf {k}'}}u}^{(1)} \xi _{{\rm {\bf
{k}''}}u}^{(1)} z_{\varphi ,{\rm {\bf k}}u - {\rm {\bf {k}'}}u - {\rm {\bf
k}}s} (\tau )}} z_{{\varphi} ',{\rm {\bf k}}s - {\rm {\bf {k}''}}u} (\tau )-
\]
\[
- {\sum\limits_{\psi ,\varphi = 1}^{2} {{\sum\limits_{{\rm {\bf {k}'}}u{\rm
{\bf {k}''}}u{\rm {\bf {k}'''}}u} {E_{\psi ,\varphi}  ({\rm {\bf k}}_{u}
,{\rm {\bf {k}'}}_{u} ,{\rm {\bf k}}_{u} - {\rm {\bf {k}'}}_{u} ,{\rm {\bf
{k}''}}_{u} ,{\rm {\bf {k}'''}}_{u} )}} \xi _{{\rm {\bf {k}'}}u}^{(1)}}
}\xi _{{\rm {\bf {k}''}}u}^{(1)} \xi _{{\rm {\bf {k}'''}}u}^{(1)} z_{\varphi
,{\rm {\bf k}}u - {\rm {\bf {k}'}}u - {\rm {\bf {k}''}}u - {\rm {\bf
{k}'''}}u} (\tau ) -
\]
\[
 - {\sum\limits_{\psi ,\varphi = 1}^{2} {{\sum\limits_{{\rm {\bf {k}'}}u{\rm
{\bf {k}''}}u{\rm {\bf {k}'''}}u} {F_{\psi ,\varphi}  ({\rm {\bf k}}_{u}
,{\rm {\bf {k}'}}_{u} ,{\rm {\bf {k}''}}_{u} ,{\rm {\bf {k}'''}}_{u} )}} \xi
_{{\rm {\bf {k}'}}u}^{(1)} \xi _{{\rm {\bf {k}''}}u}^{(1)} \xi _{{\rm {\bf
{k}'''}}u}^{(1)} z_{\varphi ,{\rm {\bf k}}u - {\rm {\bf {k}'}}u - {\rm {\bf
{k}''}}u - {\rm {\bf {k}'''}}u} (\tau )}}  -
\]
\[
 - {\sum\limits_{\psi ,{\varphi} ',\varphi = 1}^{2} {{\sum\limits_{{\rm {\bf
{k}'}}u{\rm {\bf {k}''}}u{\rm {\bf {k}'''}}u{\rm {\bf k}}s} {G_{\psi
,\varphi ,{\varphi} '} ({\rm {\bf k}}_{u} ,{\rm {\bf {k}'}}_{u} ,{\rm {\bf
k}}_{s} ,{\rm {\bf {k}''}}_{u} ,{\rm {\bf {k}'''}}_{u} )}} \xi _{{\rm {\bf
{k}'}}u}^{(1)} \xi _{{\rm {\bf {k}''}}u}^{(1)} \xi _{{\rm {\bf
{k}'''}}u}^{(1)} z_{\varphi ,{\rm {\bf k}}u - {\rm {\bf {k}'}}u - {\rm {\bf
k}}s} (\tau )z_{{\varphi} ',{\rm {\bf k}}s - {\rm {\bf {k}''}}u - {\rm {\bf
{k}'''}}u} (\tau )}} .
\]

\noindent
Function $\omega ({\rm {\bf k}}_{u} ,{\rm {\bf {k}'}}_{u} ,{\rm {\bf
{k}''}}_{u} ,{\rm {\bf {k}'''}}_{u} ,{\rm {\bf k}}_{u} - {\rm {\bf
{k}'}}_{u} )$ and others introduced in equations (\ref{eq14}) are presented in the
Appendix B. The modes $\xi _{{\rm {\bf k}}u}^{(1)} $ are order parameters.
Their collaboration or competition determines the behavior of the system.

The system of equations (\ref{eq14}) is still difficult to analyze because it
contains random components. Further analysis of the equations (\ref{eq14}) may
consist in their statistical averaging or in obtaining the Fokker-Planck
equation.

\section{Statistical averaging}

For statistical averaging we used the relationship between the moments and
the correlation functions \cite{Strat36} and multi-dimensional generalization of
Furutsu - Novikov formula \cite{Klyat31}. Taking into account the formal solution of
Eqs. (\ref{eq14}) and assuming, that we can neglect higher than second order
correlation functions, it can be shown, that in Eqs. (\ref{eq14})the terms
containing the product $\xi _{{\rm {\bf {k}'}}u}^{(1)} \xi _{{\rm {\bf
{k}''}}u}^{(1)} \xi _{{\rm {\bf {k}'''}}u}^{(1)} z_{\varphi ,{\rm {\bf
k}}_{1}}  (\tau )z_{{\varphi} ',{\rm {\bf k}}_{2}}  (\tau )$ and $\xi _{{\rm
{\bf {k}'}}u}^{(1)} \xi _{{\rm {\bf {k}''}}u}^{(1)} \xi _{{\rm {\bf
{k}'''}}u}^{(1)} z_{\varphi ,{\rm {\bf k}}} (\tau )$ must be discarded since
they do not contribute to the average values of the modes $\xi _{{\rm {\bf
k}}u}^{(1)} $ in averaging. We note here, that the procedure of correlation
splitting leads to the appearance of a similar correlation for the other
interacting modes. Therefore, this procedure should be performed until all
the terms containing the necessary degree of intensity fluctuations are
taken into account. The remaining correlation functions can be neglected
owing to their smallness, as the terms obtained after their splitting are
proportional to a higher degree of noise intensity.

In order to obtain after averaging corrections to the increments of unstable
mode amplitudes with an accuracy to terms quadratic in the noise intensity,
when calculating the functional derivatives in the Furutsu - Novikov formula
it is necessary to retain the terms containing the product $\xi _{{\rm {\bf
{k}'}}u}^{(1)} z_{\varphi ,{\rm {\bf k}}} (\tau )$.

Below is given only the structure of equations obtained by the procedure of
averaging since they have a very complicated form:

\begin{equation}
\label{eq15}
{\frac{{d{\left\langle {\xi _{{\rm {\bf k}}u}^{(1)}}  \right\rangle
}}}{{d\tau}} } - \lambda _{1} ({\rm {\bf k}}_{u} ){\left\langle {\xi _{{\rm
{\bf k}}u}^{(1)}}  \right\rangle}  = {\rm {\bf L}}_{0} \left( {{\rm {\bf
k}}_{u}}  \right) + {\rm {\bf L}}_{1} \left( {{\rm {\bf k}}_{u} ,{\rm {\bf
{k}'}}_{u}}  \right){\left\langle {\xi _{{\rm {\bf {k}'}}u}^{(1)}}
\right\rangle}  +
\end{equation}
\begin{displaymath}
 + {\rm {\bf L}}_{2} \left( {{\rm {\bf k}}_{u} ,{\rm {\bf {k}'}}_{u} ,{\rm
{\bf {k}''}}_{u} ,{\rm {\bf {k}'''}}_{u} ,{\rm {\bf k}}_{s}}
\right){\left\langle {\xi _{{\rm {\bf {k}''}}u}^{(1)} \xi _{{\rm {\bf
{k}'''}}u}^{(1)}}  \right\rangle}  + {\rm {\bf L}}_{3} \left( {{\rm {\bf
k}}_{u} ,{\rm {\bf {k}'}}_{u} ,{\rm {\bf {k}''}}_{u} ,{\rm {\bf {k}'''}}_{u}
,{\rm {\bf k}}_{u}^{IV} ,{\rm {\bf k}}_{s} ,{\rm {\bf {k}'}}_{s}}
\right){\left\langle {\xi _{{\rm {\bf {k}''}}u}^{(1)} \xi _{{\rm {\bf
{k}'''}}u}^{(1)} \xi _{{\rm {\bf k}}^{IV}u}^{(1)}}  \right\rangle} .
\end{displaymath}

The analysis of the equations (\ref{eq15}) leads to the following conclusions.

First, after the averaging of Eqs. (\ref{eq14}) additional terms, that do not depend
on ${\left\langle {\xi _{{\rm {\bf k}}u}^{(1)}}  \right\rangle} $, arise in
(\ref{eq15}). They are determined by the parameters of the problem, the type of the
correlation function $g_{jj}$, the noise intensity, and the wavenumber of a
given mode.

Second, in the system (\ref{eq15}) there are additional terms proportional to
${\left\langle {\xi _{{\rm {\bf k}}u}^{(1)}}  \right\rangle} $. This leads
to a variation of eigenvalues of unstable mode  amplitudes in
comparison with the deterministic case. As a result, the region of
instability of the system, the beginning of the process of destruction of a
statistically stationary homogeneous state and pattern formation as well as
the duration of the transitional regime from one statistically stationary
state to another are changed.

Selecting from ${\rm {\bf L}}_{1} \left( {{\rm {\bf k}}_{u} ,{\rm {\bf
{k}'}}_{u}}  \right)$ the terms that contribute to the increment of
${\left\langle {\xi _{{\rm {\bf k}}u}^{(1)}}  \right\rangle} $, we obtain
the dispersion equation

\begin{equation}
\label{eq16}
\lambda = \lambda _{1} ({\rm {\bf k}}_{u} ) + {\rm {\bf L}}_{1} \left( {{\rm
{\bf k}}_{u} ,{\rm {\bf k}}_{u}}  \right) + {\rm {\bf L}}_{1} \left( {{\rm
{\bf k}}_{u} , - {\rm {\bf k}}_{u}}  \right).
\end{equation}

When deriving eq.(\ref{eq16}) we take into account the, fact that $\xi _{ - {\rm
{\bf k}}u}^{(1)} = \xi _{{\rm {\bf k}}u}^{ * (1)} = \xi _{{\rm {\bf
k}}u}^{(1)} $, because the solutions of equations (\ref{eq14}) must be real. From
equation (\ref{eq16}) and expressions ${\rm {\bf L}}_{1} \left( {{\rm {\bf k}}_{u}
,{\rm {\bf {k}'}}_{u}}  \right)$ and $g_{ii} $, it follows, that
increments (Re $\lambda )$ of the unstable mode averaged amplitudes are proportional
to the intensity of noise and depend on the correlation length. Herewith, the
intensity of noise becomes another bifurcation parameter and patterns
begin to form, when the parameter \textit{$\chi $}$_{{\rm 2}\theta} $ is different from
\textit{$\chi $}$_{{\rm 2 c}}$. Research of particular systems shows, that the value of
the bifurcation parameter \textit{$\chi $}$_{{\rm 2}}$ is shifted to the subcritical region
for the case of deterministic description. Finally, we note, that in the
external random fields pattern formation occurs due to multimode
interactions, whereby the conditions the resonant interaction of the modes
are also different from the deterministic case. In particular, interaction
between different configurations of modes with the modes with doubled
wavenumbers arises, for example $2{\rm {\bf {k}'}}_{u} - {\rm {\bf
{k}''}}_{u} = {\rm {\bf k}}_{u} $, as well as the five-mode interaction.

Thus, the above theoretical analysis allows us to describe the evolution of
the stochastic systems under consideration, in the vicinity of the Turing bifurcation
point in more detail.

\section{Fokker-Planck equation for the order parameters}

If the system parameters are such that it is in the supercritical region
then with increasing noise intensity, the system will "go" farther and
farther from the bifurcation point of a deterministic system. To describe
the state of the system in this case we can use the Fokker-Planck equation.

For the system (\ref{eq14}) the Fokker - Planck equation can be written in general
form as follows [35]:

\begin{equation}
\label{eq17}
{\frac{{\partial w\left( {{\left\{ {\xi _{{\rm {\bf k}}u}^{(1)}}
\right\}},\tau}  \right)}}{{\partial \tau}} } = - {\sum\limits_{{\rm {\bf
k}}u} {{\frac{{\partial}} {{\partial \xi _{{\rm {\bf k}}u}^{(1)}}} }{\left\{
{\left( { \langle F_{{\rm {\bf k}}u} (\tau ) \rangle + {\sum\limits_{{\rm {\bf q}}u}
{{\int\limits_{ - \infty} ^{0} {K[{\frac{{\partial F_{{\rm {\bf k}}u} (\tau
)}}{{\partial \xi _{{\rm {\bf q}}u}^{(1)}}} },F_{{\rm {\bf q}}u} ({t}')]}}}
} d{t}'} \right)w} \right\}}}}  +
\end{equation}
\begin{displaymath}
 + {\sum\limits_{{\rm {\bf k}}u,{\rm {\bf q}}u} {{\frac{{\partial
^{2}}}{{\partial \xi _{{\rm {\bf k}}u}^{(1)} \partial \xi _{{\rm {\bf
q}}u}^{(1)}}} }{\left\{ {\left( {{\int\limits_{ - \infty} ^{0} {K[F_{{\rm
{\bf k}}u} (\tau ),F_{{\rm {\bf q}}u} ({t}')]}} d{t}'} \right)w}
\right\}}}} .
\end{displaymath}

\noindent
Here $w\left( {{\left\{ {\xi _{{\rm {\bf k}}u}^{(1)}}  \right\}},\tau}
\right)$ is multivariate probability density, which determines the
probability of some configuration of unstable modes ${\left\{ {\xi _{{\rm
{\bf k}}u}^{(1)}}  \right\}}$. After transformation with an accuracy to
terms linear in the noise intensity we can obtain the correlation functions
appearing in (\ref{eq17}):

\[
K[{\frac{{\partial F_{{\rm {\bf k}}u} (\tau )}}{{\partial \xi _{{\rm {\bf
q}}u}^{(1)}}} },F_{{\rm {\bf q}}u} ({t}')] = {\sum\limits_{\varphi}  {\eta
_{\varphi}  ({\rm {\bf k}}_{u} ,{\rm {\bf q}}_{u} )O_{\varphi} ^{ * (1)}
({\rm {\bf q}}_{u} )p_{\varphi} ^{(0)} g_{\varphi \varphi}  (\vert {\rm {\bf
k}}_{u} - {\rm {\bf q}}_{u} \vert )}} \delta _{{\rm {\bf k}}u - {\rm {\bf
q}}u,{\rm {\bf q}}u} \delta (\tau - {t}') +
\]
\[
 + {\sum\limits_{\varphi}  {[\nu _{\varphi}  ({\rm {\bf k}}_{u} ,{\rm {\bf
q}}_{u} ,{\rm {\bf k}}_{u} - 2{\rm {\bf q}}_{u} ) + \nu _{\varphi}  ({\rm
{\bf k}}_{u} ,{\rm {\bf k}}_{u} - 2{\rm {\bf q}}_{u} ,{\rm {\bf q}}_{u} )]}
}
O_{\varphi} ^{ * (1)} ({\rm {\bf q}}_{u} )p_{\varphi} ^{(0)} g_{\varphi
\varphi}  (\vert {\rm {\bf q}}_{u} \vert )\xi _{{\rm {\bf k}}u - 2{\rm {\bf
q}}u} \delta (\tau - {t}') +
\]
\[
 + {\sum\limits_{\varphi}  {\eta _{\varphi}  ({\rm {\bf k}}_{u} ,{\rm {\bf
q}}_{u} )\eta _{\varphi}  ({\rm {\bf q}}_{u} ,2{\rm {\bf q}}_{u} - {\rm {\bf
k}}_{u} )g_{\varphi \varphi}  (\vert {\rm {\bf k}}_{u} - {\rm {\bf q}}_{u}
\vert )\xi _{2{\rm {\bf q}}u - {\rm {\bf k}}u}}}  \delta (\tau - {t}') +
\]
\[
 + {\sum\limits_{\varphi ,{\rm {\bf {q}'}}u} {\eta _{\varphi}  ({\rm {\bf
k}}_{u} ,{\rm {\bf q}}_{u} )\nu _{\varphi}  ({\rm {\bf q}}_{u} ,{\rm {\bf
{q}'}}_{u} ,2{\rm {\bf q}}_{u} - {\rm {\bf k}}_{u} - {\rm {\bf {q}'}}_{u}
)g_{\varphi \varphi}  (\vert {\rm {\bf k}}_{u} - {\rm {\bf q}}_{u} \vert
)\xi _{{\rm {\bf {q}'}}u} \xi _{2{\rm {\bf q}}u - {\rm {\bf k}}u - {\rm {\bf
{q}'}}u}}}  \delta (\tau - {t}') +
\]
\[
 + {\sum\limits_{\varphi ,{\rm {\bf {q}'}}u,{\rm {\bf {q}''}}u} {[\nu
_{\varphi}  ({\rm {\bf k}}_{u} ,{\rm {\bf q}}_{u} ,{\rm {\bf k}}_{u} - 2{\rm
{\bf q}}_{u} + {\rm {\bf {q}'}}_{u} + {\rm {\bf {q}''}}_{u} ) + \nu
_{\varphi}  ({\rm {\bf k}}_{u} ,{\rm {\bf k}}_{u} - 2{\rm {\bf q}}_{u} +
{\rm {\bf {q}'}}_{u} + {\rm {\bf {q}''}}_{u} ,{\rm {\bf q}}_{u} )]\times}}
\]
\[
\times \nu _{\varphi}  ({\rm {\bf q}}_{u} ,{\rm {\bf {q}'}}_{u} ,{\rm {\bf
{q}''}}_{u} )g_{\varphi \varphi}  (\vert {\rm {\bf q}}_{u} - {\rm {\bf
{q}'}}_{u} - {\rm {\bf {q}''}}_{u} \vert )\xi _{{\rm {\bf {q}'}}u} \xi
_{{\rm {\bf {q}''}}u} \xi _{{\rm {\bf k}}u - 2{\rm {\bf q}}u + {\rm {\bf
{q}'}}u + {\rm {\bf {q}''}}u} \delta (\tau - {t}') +
\]
\[
+ {\sum\limits_{\varphi ,{\rm {\bf {q}'}}u} {\eta _{\varphi}  ({\rm {\bf
q}}_{u} ,{\rm {\bf {q}'}}_{u} )[\nu _{\varphi}  ({\rm {\bf k}}_{u} ,{\rm
{\bf k}}_{u} - 2{\rm {\bf q}}_{u} + {\rm {\bf {q}'}}_{u} ,{\rm {\bf q}}_{u}
) + \nu _{\varphi}  ({\rm {\bf k}}_{u} ,{\rm {\bf q}}_{u} ,{\rm {\bf k}}_{u}
- 2{\rm {\bf q}}_{u} + {\rm {\bf {q}'}}_{u} )]}} \times
\]
\[
\times g_{\varphi \varphi}  (\vert {\rm {\bf q}}_{u} - {\rm {\bf {q}'}}_{u}
\vert )\xi _{{\rm {\bf {q}'}}u} \xi _{{\rm {\bf k}}u - 2{\rm {\bf q}}u +
{\rm {\bf {q}'}}u} \delta (\tau - {t}'),
\]

\[
K[F_{{\rm {\bf k}}u} (\tau ),F_{{\rm {\bf q}}u} ({t}')] =
{\sum\limits_{\varphi}  {{\left[ {O_{\varphi} ^{ * (1)} ({\rm {\bf k}}_{u}
)} \right]}^{2}\left( {p_{\varphi} ^{(0)}}  \right)^{2}g_{\varphi \varphi}
(\vert {\rm {\bf k}}_{u} \vert )}} \delta _{{\rm {\bf k}}u,{\rm {\bf q}}u}
\delta (\tau - {t}') +
\]
\[
 + {\sum\limits_{\varphi}  {\eta _{\varphi}  ({\rm {\bf k}}_{u} ,{\rm {\bf
k}}_{u} - {\rm {\bf q}}_{u} )O_{\varphi} ^{ * (1)} ({\rm {\bf q}}_{u}
)p_{\varphi} ^{(0)} g_{\varphi \varphi}  (\vert {\rm {\bf q}}_{u} \vert )\xi
_{{\rm {\bf k}}u - {\rm {\bf q}}u}}}  \delta (\tau - {t}') +
\]
\[
 + {\sum\limits_{\varphi ,{\rm {\bf {k}'u}}} {\nu _{\varphi}  ({\rm {\bf
k}}_{u} ,{\rm {\bf {k}'}}_{u} ,{\rm {\bf k}}_{u} - {\rm {\bf {k}'}}_{u} -
{\rm {\bf q}}_{u} )O_{\varphi} ^{ * (1)} ({\rm {\bf q}}_{u} )p_{\varphi
}^{(0)} g_{\varphi \varphi}  (\vert {\rm {\bf q}}_{u} \vert )\xi _{{\rm {\bf
{k}'}}u}}}  \xi _{{\rm {\bf k}}u - {\rm {\bf {k}'}}u - {\rm {\bf q}}u}
\delta (\tau - {t}') +
\]
\[
 + {\sum\limits_{\varphi}  {\eta _{\varphi}  ({\rm {\bf q}}_{u} ,{\rm {\bf
q}}_{u} - {\rm {\bf k}}_{u} )O_{\varphi} ^{ * (1)} ({\rm {\bf k}}_{u}
)p_{\varphi} ^{(0)} g_{\varphi \varphi}  (\vert {\rm {\bf k}}_{u} \vert )\xi
_{{\rm {\bf q}}u - {\rm {\bf k}}u}}}  \delta (\tau - {t}') +
\]
\[
 + {\sum\limits_{\varphi ,{\rm {\bf {q}'u}}} {\eta _{\varphi}  ({\rm {\bf
k}}_{u} ,{\rm {\bf k}}_{u} - {\rm {\bf q}}_{u} + {\rm {\bf {q}'}}_{u} )\eta
_{\varphi}  ({\rm {\bf q}}_{u} ,{\rm {\bf {q}'}}_{u} )g_{\varphi \varphi}
(\vert {\rm {\bf q}}_{u} - {\rm {\bf {q}'}}_{u} \vert )\xi _{{\rm {\bf
{q}'}}u}}}  \xi _{{\rm {\bf k}}u - {\rm {\bf q}}u + {\rm {\bf {q}'}}u}
\delta (\tau - {t}') +
\]
\[
 + {\sum\limits_{\varphi ,{\rm {\bf {k}'u}},{\rm {\bf {q}'u}}} {\nu _{\varphi
} ({\rm {\bf k}}_{u} ,{\rm {\bf {k}'}}_{u} ,{\rm {\bf k}}_{u} - {\rm {\bf
{k}'}}_{u} - {\rm {\bf q}}_{u} + {\rm {\bf {q}'}}_{u} )\eta _{\varphi}
({\rm {\bf q}}_{u} ,{\rm {\bf {q}'}}_{u} )}} g_{\varphi \varphi}  (\vert
{\rm {\bf q}}_{u} - {\rm {\bf {q}'}}_{u} \vert )\xi _{{\rm {\bf {k}'}}u} \xi
_{{\rm {\bf {q}'}}u} \xi _{{\rm {\bf k}}u - {\rm {\bf {k}'}}u - {\rm {\bf
q}}u + {\rm {\bf {q}'}}u} \delta (\tau - {t}') +
\]
\[
 + {\sum\limits_{\varphi ,{\rm {\bf {q}'u}}} {\nu _{\varphi}  ({\rm {\bf
q}}_{u} ,{\rm {\bf {q}'}}_{u} ,{\rm {\bf q}}_{u} - {\rm {\bf {q}'}}_{u} -
{\rm {\bf k}}_{u} )O_{\varphi} ^{ * (1)} ({\rm {\bf k}}_{u} )p_{\varphi
}^{(0)} g_{\varphi \varphi}  (\vert {\rm {\bf k}}_{u} \vert )\xi _{{\rm {\bf
{q}'}}u}}}  \xi _{{\rm {\bf q}}u - {\rm {\bf {q}'}}u - {\rm {\bf k}}u}
\delta (\tau - {t}') +
\]
\[
 + {\sum\limits_{\varphi ,{\rm {\bf {k}'u}},{\rm {\bf {q}'u}}} {\nu _{\varphi
} ({\rm {\bf q}}_{u} ,{\rm {\bf {q}'}}_{u} ,{\rm {\bf q}}_{u} - {\rm {\bf
{q}'}}_{u} - {\rm {\bf k}}_{u} + {\rm {\bf {k}'}}_{u} )\eta _{\varphi}
({\rm {\bf k}}_{u} ,{\rm {\bf {k}'}}_{u} )}} g_{\varphi \varphi}  (\vert
{\rm {\bf k}}_{u} - {\rm {\bf {k}'}}_{u} \vert )\xi _{{\rm {\bf {k}'}}u} \xi
_{{\rm {\bf {q}'}}u} \xi _{{\rm {\bf q}}u - {\rm {\bf {q}'}}u - {\rm {\bf
k}}u + {\rm {\bf {k}'}}u} \delta (\tau - {t}') +
\]
\[
 + {\sum\limits_{\varphi ,{\rm {\bf {k}'u}},{\rm {\bf {q}'u}},{\rm {\bf
{q}''u}}} {\nu _{\varphi}  ({\rm {\bf k}}_{u} ,{\rm {\bf {k}'}}_{u} ,{\rm
{\bf k}}_{u} - {\rm {\bf {k}'}}_{u} - {\rm {\bf q}}_{u} + {\rm {\bf
{q}'}}_{u} + {\rm {\bf {q}''}}_{u} )\nu _{\varphi}  ({\rm {\bf q}}_{u} ,{\rm
{\bf {q}'}}_{u} ,{\rm {\bf {q}''}}_{u} )\times}}
\]
\[
\times g_{\varphi \varphi}  (\vert
{\rm {\bf q}}_{u} - {\rm {\bf {q}'}}_{u} - {\rm {\bf {q}''}}_{u} \vert )\xi
_{{\rm {\bf {k}'}}u} \xi _{{\rm {\bf {q}'}}u} \xi _{{\rm {\bf {q}''}}u} \xi
_{{\rm {\bf k}}u - {\rm {\bf {k}'}}u - {\rm {\bf q}}u + {\rm {\bf {q}'}}u +
{\rm {\bf {q}''}}u} \delta (\tau - {t}').
\]

Suppose, that the space of the system under study is two-dimensional. If in
this space there is only one unstable mode with the wave vector ${\rm {\bf
k}}_{c} $ and amplitude $\xi _{{\rm {\bf k}}c} $ the equation (\ref{eq17}) is
significantly simplified

\begin{equation}
\label{eq18}
{\frac{{\partial w(\xi _{{\rm {\bf k}}c} ,\tau )}}{{\partial \tau}} } = -
{\frac{{\partial}} {{\partial \xi _{{\rm {\bf k}}c}}} }{\left[ {(a\xi _{{\rm
{\bf k}}c} + b\xi _{{\rm {\bf k}}c}^{3} )w - (c + d\xi _{{\rm {\bf k}}c}^{2}
+ e\xi _{{\rm {\bf k}}c}^{4} ){\frac{{\partial w}}{{\partial \xi _{{\rm {\bf
k}}c}}} }} \right]}.
\end{equation}

Here

\[
a = \lambda _{1} ({\rm {\bf k}}_{c} ) -
{\frac{{1}}{{2}}}{\sum\limits_{\varphi}  {\eta _{\varphi} ^{2} ({\rm {\bf
k}}_{c} ,{\rm {\bf k}}_{c} )}} g_{\varphi \varphi}  (0) -
{\sum\limits_{\varphi}  {\nu _{\varphi}  ({\rm {\bf k}}_{c} ,{\rm {\bf
k}}_{c} ,{\rm {\bf k}}_{c} )O_{\varphi} ^{ * (1)} ({\rm {\bf k}}_{c}
)p_{\varphi} ^{(0)} g_{\varphi \varphi}  ({\rm {\bf k}}_{c} )}} ,
\]

\[
b = \omega ({\rm {\bf k}}_{c} ,{\rm {\bf k}}_{c} ,{\rm {\bf k}}_{c} ,{\rm
{\bf k}}_{c} ,0) - {\sum\limits_{\varphi}  {\nu _{\varphi} ^{2} ({\rm {\bf
k}}_{c} ,{\rm {\bf k}}_{c} ,{\rm {\bf k}}_{c} )g_{\varphi \varphi}  ({\rm
{\bf k}}_{c} )}} ,
\]

\[
c ={\frac{{1}}{{2}}}{\sum\limits_{\varphi}  {{\left[ {O_{\varphi} ^{\ast
(1)} ({\rm {\bf k}}_{c} )p_{\varphi} ^{(0)}}  \right]}^{2}g_{\varphi \varphi
} ({\rm {\bf k}}_{c} )}} ,
\]

\[
d = {\frac{{1}}{{2}}}{\sum\limits_{\varphi}  {\eta _{\varphi} ^{2} ({\rm
{\bf k}}_{c} ,{\rm {\bf k}}_{c} )}} g_{\varphi \varphi}  (0) +
{\sum\limits_{\varphi}  {\nu _{\varphi}  ({\rm {\bf k}}_{c} ,{\rm {\bf
k}}_{c} ,{\rm {\bf k}}_{c} )O_{\varphi} ^{ * (1)} ({\rm {\bf k}}_{c}
)p_{\varphi} ^{(0)} g_{\varphi \varphi}  ({\rm {\bf k}}_{c} )}} ,
\]

\[
e = {\frac{{1}}{{2}}}{\sum\limits_{\varphi}  {\nu _{\varphi} ^{2} ({\rm {\bf
k}}_{c} ,{\rm {\bf k}}_{c} ,{\rm {\bf k}}_{c} )g_{\varphi \varphi}  ({\rm
{\bf k}}_{c} )}} .
\]

The stationary solution of equation (\ref{eq18}) has the form:

\begin{equation}
\label{eq19}
\begin{array}{l}
 w(\xi _{{\rm {\bf k}}c} ) = N{\left| {c + d\xi _{{\rm {\bf k}}c}^{2} + e\xi
_{{\rm {\bf k}}c}^{4}}  \right|}^{{\frac{{b}}{{4e}}}}{\left| {{\frac{{2e\xi
_{{\rm {\bf k}}c}^{2} + d - \sqrt {d^{2} - 4ec}}} {{2e\xi _{{\rm {\bf
k}}c}^{2} + d + \sqrt {d^{2} - 4ec}}} }} \right|}^{{\frac{{2ae -
bd}}{{4e\sqrt {d^{2} - 4ec}}} }},\,\,\,\,\,\,\,\,\,\,\,\,\,\,\,\,\,\,d^{2} >
4ec, \\
 \,\,\,\,\,\,\,\,\,\,\,\,\,\,\,\, = N{\left| {c + d\xi _{{\rm {\bf k}}c}^{2}
+ e\xi _{{\rm {\bf k}}c}^{4}}  \right|}^{{\frac{{b}}{{4e}}}}\exp {\left[
{{\frac{{2ae - bd}}{{2e\sqrt {4ec - d^{2}}}} }{\rm a}{\rm r}{\rm c}{\rm
t}{\rm g}\left( {{\frac{{2e\xi _{{\rm {\bf k}}c}^{2} + d}}{{\sqrt {4ec -
d^{2}}}} }} \right)} \right]},\,\,4ec > d^{2}. \\
 \end{array}
\end{equation}

Here $N$ is the normalization constant:

\begin{equation}
\label{eq20}
\begin{array}{l}
 N = 1 / {\int\limits_{ - \infty} ^{ + \infty}  {{\left| {c + d\xi _{{\rm
{\bf k}}c}^{2} + e\xi _{{\rm {\bf k}}c}^{4}}
\right|}^{{\frac{{b}}{{4e}}}}{\left| {{\frac{{2e\xi _{{\rm {\bf k}}c}^{2} +
d - \sqrt {d^{2} - 4ec}}} {{2e\xi _{{\rm {\bf k}}c}^{2} + d + \sqrt {d^{2} -
4ec}}} }} \right|}^{{\frac{{2ae - bd}}{{4e\sqrt {d^{2} - 4ec}}} }}d\xi
_{{\rm {\bf k}}c}}}  ,\,\,\,\,\,\,\,\,\,\,\,\,\,\,\,\,\,d^{2} > 4ec, \\
 \,\,\,\,\,\,\, = 1 / {\int\limits_{ - \infty} ^{ + \infty}  {{\left| {c +
d\xi _{{\rm {\bf k}}c}^{2} + e\xi _{{\rm {\bf k}}c}^{4}}
\right|}^{{\frac{{b}}{{4e}}}}\exp {\left[ {{\frac{{2ae - bd}}{{2e\sqrt {4ec
- d^{2}}}} }{\rm a}{\rm r}{\rm c}{\rm t}{\rm g}\left( {{\frac{{2e\xi _{{\rm
{\bf k}}c}^{2} + d}}{{\sqrt {4ec - d^{2}}}} }} \right)} \right]}d\xi _{{\rm
{\bf k}}c}}}  ,\,\,4ec > d^{2}. \\
 \end{array}
\end{equation}

In the following section noise-induced effects, which arise during Turing
pattern formation in the well-known biophysical system, will be studied. We
compare the analytical results obtained using the above-developed approach
with the results of numerical experiments.

\section{Noise-induced effects in one complicated biophysical system}

A mathematical model of the system under study is described by equations
\cite{Schef37,Malchow38,Malchow39}:

\begin{equation}
\label{eq21}
\begin{array}{l}
 {\frac{{\partial x_{1}}} {{\partial t}}} = rx_{1} (1 - x_{1} ) -
{\frac{{ax_{1}x_{2}}} {{1 + bx_{1}}} } + D_{1} \nabla ^{2}x_{1} , \\
 {\frac{{\partial x_{2}}} {{\partial t}}} = {\frac{{ax_{1}x_{2}}} {{1 + bx_{1}
}}} - mx_{2} - {\frac{{g^{2}x_{2}^{2}}} {{1 + h^{2}x_{2}^{2}}} }f +
D_{2} \nabla ^{2}x_{2} , \\
 \end{array}
\end{equation}

\noindent
where $x_{{\rm 1}}$, $x_{{\rm 2}}$ are state functions, parameters \textit{r, a, b, m, g, h, f, D}$_{{\rm
1}}$, and D$_{{\rm 2}}$ are described in detail in \cite{Schef37,Malchow39}.The investigation of
local dynamics and bifurcation analysis of system (\ref{eq21}) is carried out in
papers \cite{Malchow38,Malchow39}.

We introduce the dimensionless time $\tau = rt$ and coordinates ${\rm {\bf
{x}'}} = {\rm {\bf x}}\sqrt {r / D_{1}}  $, and represent the parameters
$m/r$ and $a/r$ in the form of: $m / r = (m_{0} / r_{0} )(1 + f_{1} ({\rm {\bf
{x}'}},\tau ))$, $a / r = (a_{0} / r_{0} )(1 + f_{2} ({\rm {\bf {x}'}},\tau
))$. Here $m_{{\rm 0}}$, $r_{{\rm 0}}$, $a_{{\rm 0}}$ are spatio-temporal
averages of the corresponding parameters, random homogeneous isotropic
fields $f_{i} ({\rm {\bf {x}'}},\tau )$ determine the spatio-temporal
Gaussian fluctuations of these parameters and have zero means and
correlation functions of the form (\ref{eq3}), (\ref{eq10}). Taking into account the noise
we obtain

\begin{equation}
\label{eq22}
\begin{array}{l}
 {\frac{{\partial x_{1}}} {{\partial \tau}} } = x_{1} (1 - x_{1} ) - {\frac{{a_{0}}} {{r_{0}}} }(1 + f_{2} ({\rm {\bf {x}'}},\tau )){\frac{{x_{1}x_{2}}} {{1 + bx_{1}}} } + D_{1} {\nabla} '^{2}x_{1}, \\
 {\frac{{\partial x_{2}}} {{\partial \tau}} } = {\frac{{a_{0}}} {{r_{0}}} }(1 + f_{2} ({\rm {\bf {x}'}},\tau )){\frac{{x_{1}x_{2}}} {{1 + bx_{1}}}} - {\frac{{m_{0}}} {{r_{0}}} }(1 + f_{1} ({\rm {\bf {x}'}},\tau ))x_{2} - {\frac{{g^{2}x_{2}^{2}}} {{1 + h^{2}x_{2}^{2}}} }f +
D_{2} {\nabla} '^{2}x_{2}  , \\
 \end{array}
\end{equation}

\subsection{Analytical research}

In this section the results of the analytical investigation of system (22)
obtained on the basis of equation (\ref{eq15}), (\ref{eq16}), (\ref{eq18}) -- (\ref{eq20}) are given.

For system (\ref{eq22}) a dispersion equation for the unstable mode averaged
amplitudes of the form (\ref{eq15}) was obtained. Dependencies of real parts of
eigenvalue $\lambda $ of ${\left\langle {\xi _{{\rm {\bf k}}u}^{(1)}}
\right\rangle} $ on the wavenumbers are shown in Fig. 1.

\begin{figure}
\centering
\includegraphics[width=3.4in]{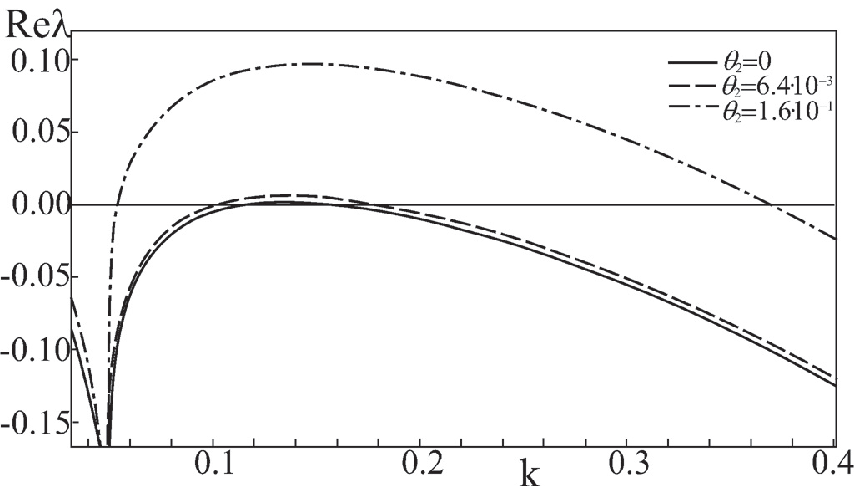}\\
(a)\\
\includegraphics[width=3.4in]{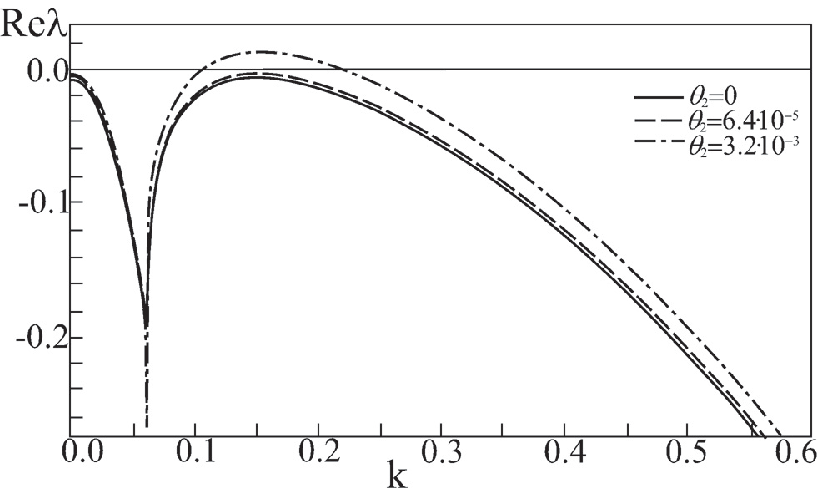}\\
(b)
\caption{ Real parts of eigenvalue \textit{$\lambda $} of unstable mode averaged amplitudes
${\left\langle {\xi _{{\rm {\bf k}}u}^{(1)}}  \right\rangle} $ versus
wavenumbers $k$ for system (22) when the noise intensity \textit{$\theta $}$_{{\rm 2}}$ changes.
(a) Supercritical region ($D$=150). (b) Subcritical region ($D$=100). For the
comparison, the figure shows curves for \textit{$\theta $}$_{{\rm 1}}$=\textit{$\theta $}$_{{\rm 2}}$=0 (solid
line). Other parameters of the model are $r_{{\rm 0}}$=1, $a_{{\rm 0}}$=8,
$g$=1.434, $f$=0.093, $h$=0.857, $b$=11.905, $ m_{{\rm 0}}$=0.490, $r_{f{\rm
1}}=r_{f{\rm 2}}$=1,\textit{ $\theta $}$_{{\rm 1}}$=2.4$\cdot $10$^{{\rm -} {\rm 5}}$. The
critical control parameter is $D$=135. }
\end{figure}

\begin{figure}
\centering
\includegraphics[width=3.4in]{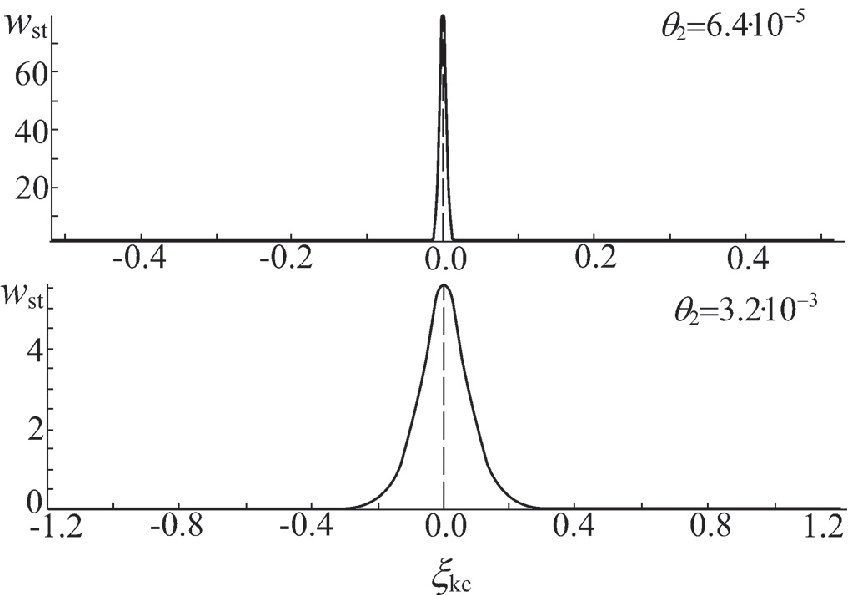}\\
\caption{Steady-state probability density (\ref{eq19}) for critical mode values of
system (\ref{eq22}) in the subcritical region for two values of noise intensity
\textit{$\theta $}$_{{\rm 2}}$. The system parameters are as in Fig.1(a).}
\end{figure}

\begin{figure}
\centering
\includegraphics[width=3.4in]{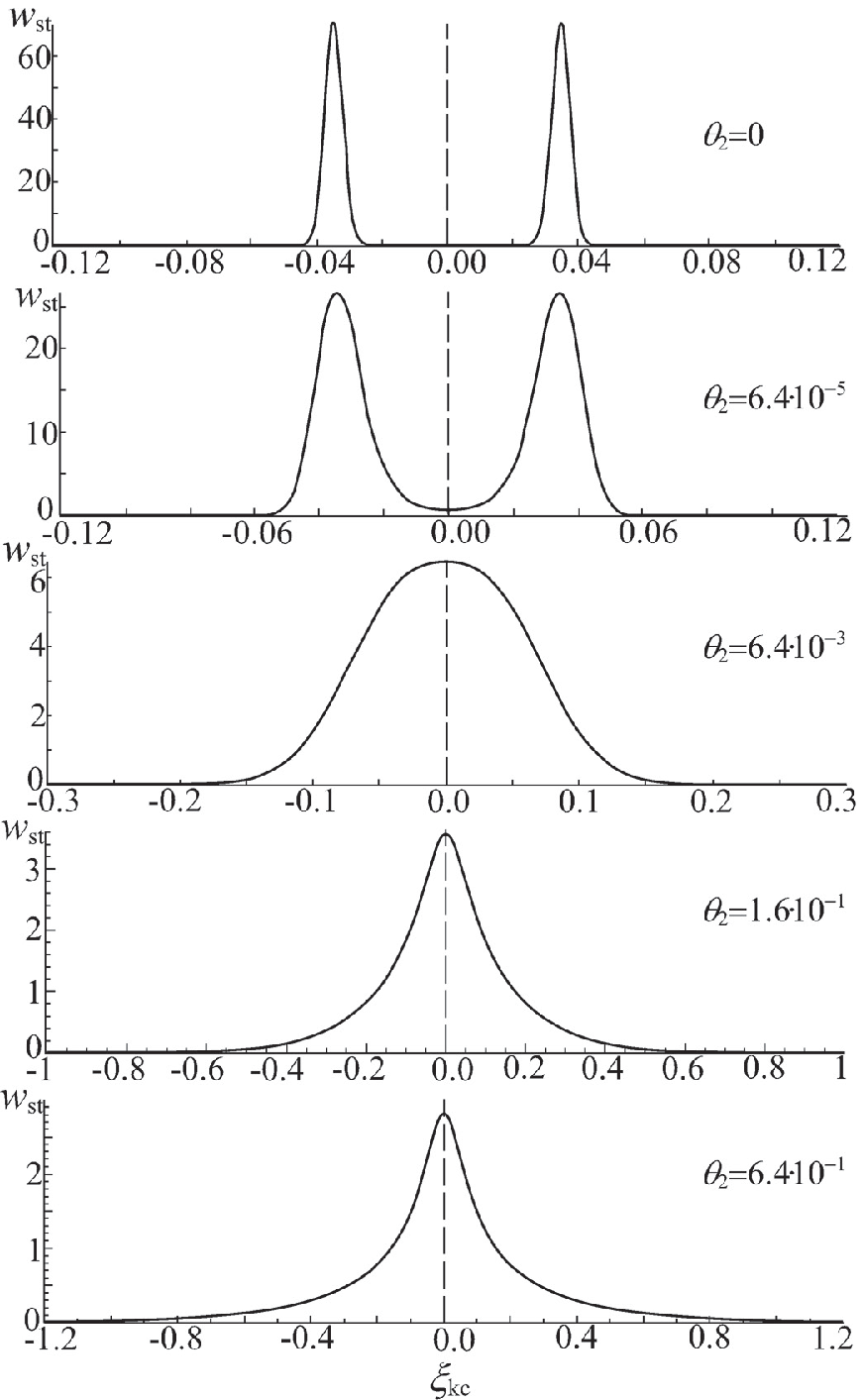}\\
\caption{Steady-state probability density (\ref{eq19}) for critical mode values of
system (\ref{eq22}) in the supercritical region for five values of noise intensity
\textit{$\theta $}$_{{\rm 2}}$. The system parameters are as in Fig.1(b).}
\end{figure}

\begin{figure}
\centering
\includegraphics[width=3.4in]{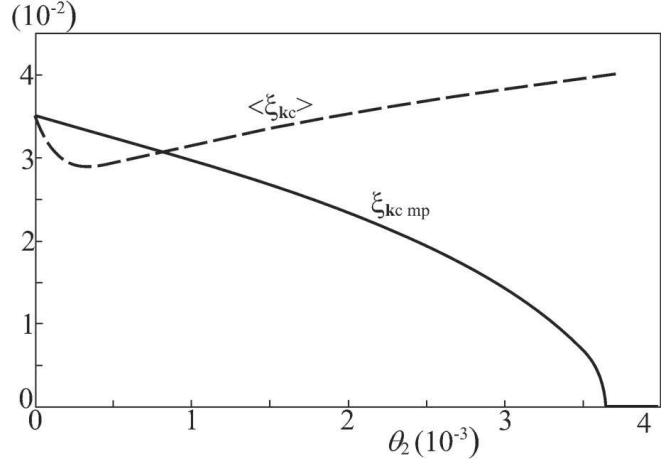}\\
\caption{The steady-state mean $\xi _{{\rm {\bf k}}c\,\,{\rm m}{\rm p}} $ and
most probable $\xi _{{\rm {\bf k}}c\,\,{\rm m}{\rm p}} $ values of the
critical mode as a function of noise intensity are given by Eqs. (\ref{eq18})-(\ref{eq20}).
Supercritical region.}
\end{figure}

\begin{figure}
\centering
\includegraphics[width=3.4in]{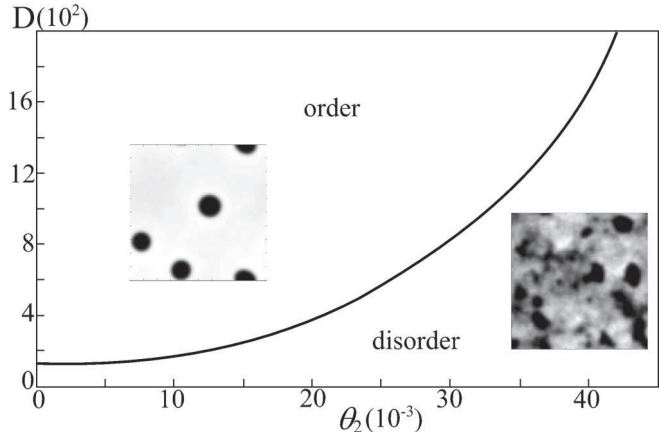}\\
\caption{The boundary of the ``order-disorder'' transition on the plane of
parameters D and $\theta _{{\rm 2}}$.}
\end{figure}

\begin{figure}
\centering
\includegraphics[width=3.4in]{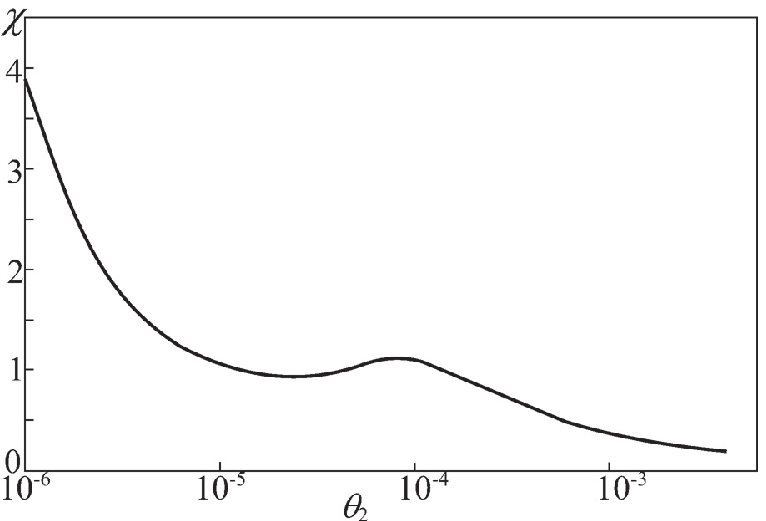}\\
(a)\\
\includegraphics[width=3.4in]{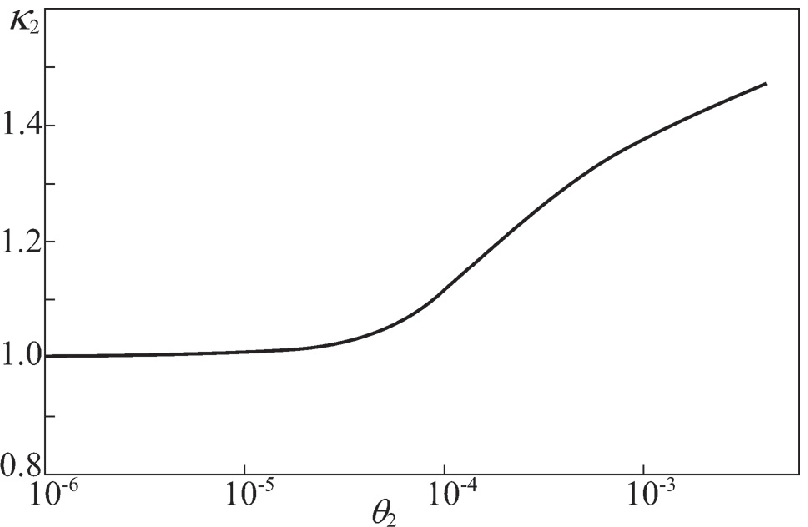}\\
(b)
\caption{ Susceptibility $\chi$ (a) and second-order
cumulant $\kappa_{2}$ (b) as functions of noise intensity \textit{ $\theta $}$_{{\rm 2}}$.}
\end{figure}

It is evident from the dependencies in Fig.1(a) (dotted and dash-dot lines),
that the region of instability of the system (\ref{eq22}) with $Re\lambda > 0$
increases with the intensity of external noise \textit{$\theta $}$_{{\rm 2}}$ in the
supercritical region. The absolute values of the increments of unstable mode
averaged amplitudes are greater in the presence of noise than in its absence.
Therefore, unstable mode amplitudes increase, on the average, considerably
faster in the presence of external noise than in its absence, which should
accelerate the process of spatial pattern formation as compared to the
deterministic description. In other words, the destruction of the
homogeneous state and pattern formation should take place earlier in time.
Besides, the expansion of the instability region and, consequently,
increasing the number of interacting unstable modes should result in the
changes of the form of the patterns.

Fig. 1(b) shows the dependencies $Re\lambda (k)$ in the subcritical region.
It is obvious, that instability does not appear in the absence of noise
(solid line in Fig. 1(b)) or in case of its low intensity (dotted line in
Fig. 1(b)). However, starting with certain intensity (dash-dot line in Fig.
1(b)) there appears a region with $Re\lambda > 0$, i.e., there exists
critical intensity of noise, that induced parametric instability of system
(22). Phase transition with pattern formation in the presence of noise
should take place in the subcritical region (i.e., earlier by the value of
the control parameter than in the case of the deterministic description).

In the subcritical region the state of system is homogeneous and
statistically stationary (disorder). Unimodal probability density is known
to correspond to this state. The appearance of inhomogeneous statistically
stationary state (order) manifests itself in the splitting of probability
density maximum into two symmetrical ones.

The Fokker - Planck equation for the critical mode probability density of
the form (\ref{eq18}) has been derived for the system (\ref{eq22}). Figures 2 and 3 show the
variation of stationary probability density of this mode with increasing
noise intensity when passing through the bifurcation point of a
deterministic system.

Fig. 2 illustrates the probability density of critical modes in the
subcritical region. If the noise is low the stationary probability density
is close to the $\delta $-function; the average and most probable values of
the critical mode coincide and are equal to zero, i.e., the homogeneous
statistically stationary state of system is the most probable one (see
top Fig.2). Deformation of the curve of stationary probability density
occurs if the noise intensity increases (see bottom Fig.2): the maximum is
considerable reduced without being shifted; the base of the curve herewith
is expanded. As the probability distribution obtained is not Gaussian the
mean value becomes different from the most probable one. Thus, although the
unimodal probability density is not split into bimodal density the mean
value of the order parameter becomes different from zero and we should
expect the inhomogeneous statistically stationary state (order) to arise,
the probability of which is not great. It is obvious, that the occurrence of
a new state is accidental in the case under consideration. It can be
explained in the following way. Formation of patterns in the subcritical
region may occur on the strong inhomogeneities of medium. Such inhomogeneities
may be produced by strong (large-scale) fluctuations, the probability of
which is not great and depends on the noise parameters. Thus, given
prolonged observation of the system's evolution and proper parameters of
external noise we might expect the formation of such random inhomogeneities,
that will cause patterns to occur in the area of their location.

Fig. 3 demonstrates the steady-state probability density of the critical
mode (\ref{eq19}) of system (\ref{eq22}) in the supercritical region at various noise
intensity. Bimodal probability densities correspond to the existence of
spatial patterns. Herewith, the most probable value and the expectation of
the order parameter become different from zero. Fig. 3 clearly shows, that
maximums gradually merge as the noise intensity increases, and unimodal
density reappears at some critical value of noise intensity. At the same
time $\xi _{{\rm {\bf k}}c\,{\rm m}{\rm p}} = 0$, and ${\left\langle
{\xi _{{\rm {\bf k}}c}}  \right\rangle}  \ne 0$. System (\ref{eq22}) transforms to
the state of strongly irregular behavior (disorder). Thus, the obtained
variation of steady-state probability density of critical mode
testifies, that there is a phase transition ``disorder - order - disorder'' in
system (\ref{eq22}).

The above-mentioned variation of the statistically stationary mean and most
probable values of the critical mode corresponding to the variation of
densities shown in Fig. 3 is illustrated in Fig. 4.

Fig. 5 presents the boundary of noise-induced phase transition
``order-disorder'' for system (\ref{eq22}), that is predicted using the approach
developed above. It should be noted, that even very small fluctuations will
contribute to the loss of stability of the inhomogeneous state and bring
about a disordered state when we approaching to the deterministic transition
point.

Let us define the relative fluctuations of the order parameter
(susceptibility) as $\chi = {{{\left[ {{\left\langle {\xi _{{\rm {\bf
k}}c}^{2}}  \right\rangle}  - {\left\langle {\xi _{{\rm {\bf k}}c}}
\right\rangle} ^{2}} \right]}} \mathord{\left/ {\vphantom {{{\left[
{{\left\langle {\xi _{{\rm {\bf k}}c}^{2}}  \right\rangle}  - {\left\langle
{\xi _{{\rm {\bf k}}c}}  \right\rangle} ^{2}} \right]}} {\theta _{2}}} }
\right. \kern-\nulldelimiterspace} {\theta _{2}}} $ and its second-order
cumulant $\kappa _{2} = {{{\left\langle {\xi _{{\rm {\bf k}}c}^{2}}
\right\rangle}}  \mathord{\left/ {\vphantom {{{\left\langle {\xi _{{\rm {\bf
k}}c}^{2}}  \right\rangle}}  {{\left\langle {\xi _{{\rm {\bf k}}c}}
\right\rangle} ^{2}}}} \right. \kern-\nulldelimiterspace} {{\left\langle
{\xi _{{\rm {\bf k}}c}}  \right\rangle} ^{2}}}$ in accordance with \cite{BroPar12}. In
Fig. 6 shows plots of $\chi $ and $\kappa _{2} $ as noise
intensity functions. The presence of susceptibility maxima (see Fig. 6(a)) clearly
shows the increase of fluctuations in the vicinity of two critical points.
The form of the curves $\kappa _{2} $ (see Fig. 6(b)) and $\chi $ obtained
for system (\ref{eq22}) coincides qualitatively in the corresponding region with the
similar curves obtained numerically in \cite{BroPar12} for system, in which a pure
noise-induced transition is observed. This testifies, that the noise-induced
transition ``order-disorder'' is of the same nature in system (\ref{eq22}).

\subsection{Simulation}

This section presents the results of simulation of the evolution of system
(\ref{eq22}) in two-dimensional space.

The following set of parameters $m_{0} / r_{0} =  0.490$, $a_{0} /
r_{0} =  8$, $g^{2} / r_{0} = {\rm 2}.{\rm 0}{\rm 5}{\rm
6}$, $f = {\rm 0}.093$, $h = {\rm 0}.857$, $b = 11.905$, $D_{2} / D_{1} =
150$, periodic boundary conditions and a rectangular domain of integration
are chosen to model the evolution of system (22) in the vicinity of the
Turing bifurcation point. Three plane waves with a low amplitude and
critical wavenumber propagated at an angle of 60$^{o}$ to each other
destabilized the spatial homogeneous state and provided hexagonal symmetry
of the initial state. The size of the integration domain is 210.0 $\times $
186.8.

The fluctuations of the parameters $m/r$ and $a/r$ are modeled as homogeneous
isotropic Gaussian fields with the zero mean and correlation functions of
the form (\ref{eq3}), (\ref{eq11}), where r$_{{\rm f}{\rm i}}$=1, k$_{{\rm t}{\rm i}}$=100.
The values of k$_{{\rm t}{\rm i}}$ inverse to the correlation time are
chosen so, that all characteristic times of a deterministic problem are
considerable greater than the correlation time. The choice of such values of
k$_{{\rm t}{\rm i}}$ ensures, that the appropriate condition is complied with
in theory.

The process of spatial pattern formation in the vicinity of the Turing
bifurcation point with increasing noise intensity $\theta _{{\rm 2}}$ is
presented in Fig. 7. In addition to the typical pattern of the state
function $x_{{\rm 1}}$ distribution over the surface the Fig. 7 also shows
the top view giving a more visual representation of the structures
configuration. To provide the top view the colour gradient from black to
white visualized the variation of values of $x_{{\rm 1}}$ from minimum to
maximum, respectively. Black areas (``cavities'') correspond to empty regions
of space. Fig. 7 shows, that in the presence of noise patterns begin to form
earlier than in the deterministic case. Besides, the greater the noise
intensity, the earlier the process of formation starts. It should be noted,
that noise destroys the symmetry of spatial patterns.

\begin{figure}
\centering
\includegraphics[width=6.62in]{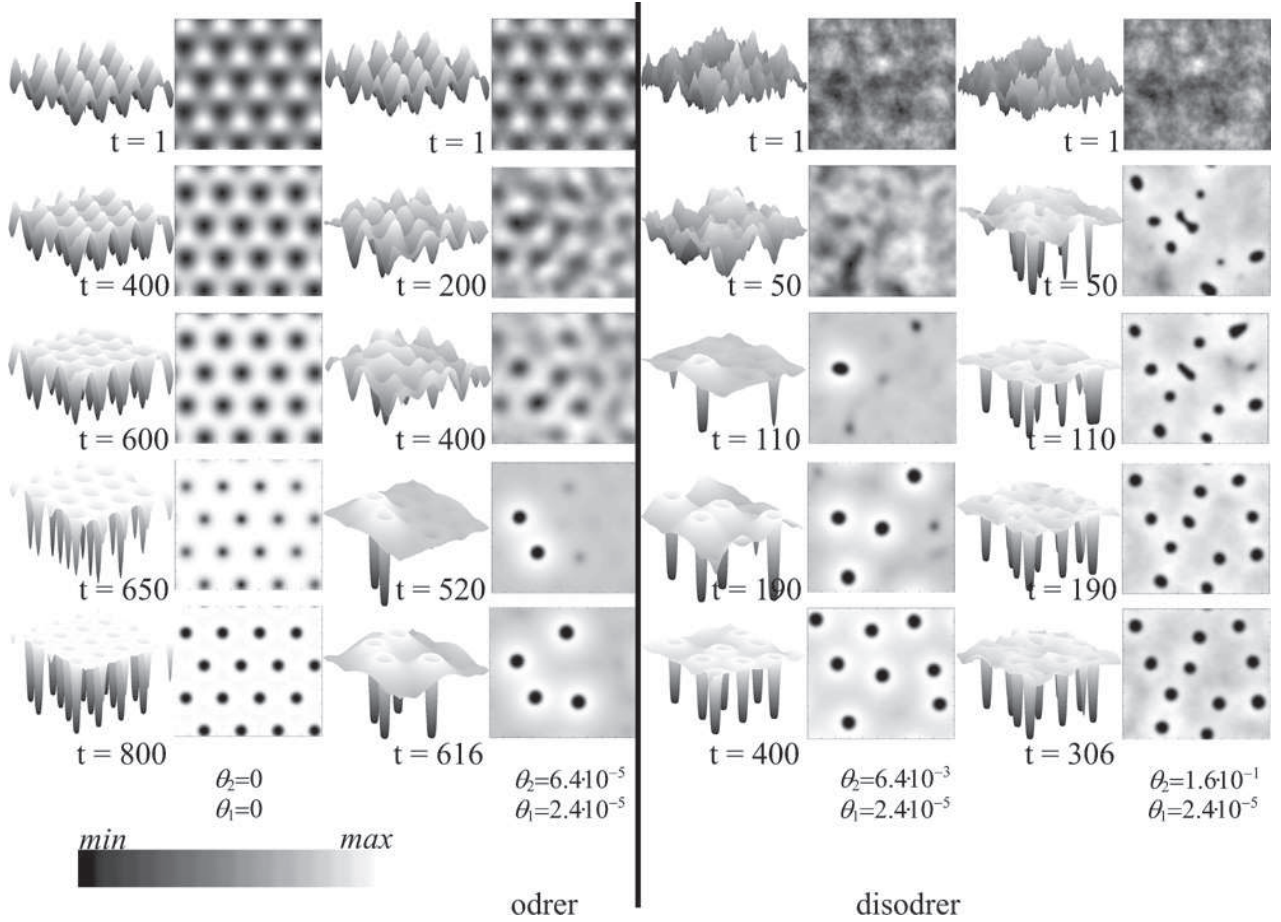}\\
\caption{The evolution of spatial Turing patterns at different noise
intensities near the deterministic transition point. On the figure the
moments of computer time corresponding to a given distribution of the
dynamic variable $x_{{\rm 1}}$ of system (\ref{eq22}) are indicated. The bottom row
of images shows statistically steady state in the case of order.}
\end{figure}

From Fig. 7 we notice, that beginning with certain noise intensity the
patterns formed become unstable; alternation of various random pattern
configurations takes place; contours of certain ``cavities'' change, i.e.,
the system transforms into the state of irregular behavior -- disorder.

\begin{figure}
\centering
\includegraphics[width=6.50in]{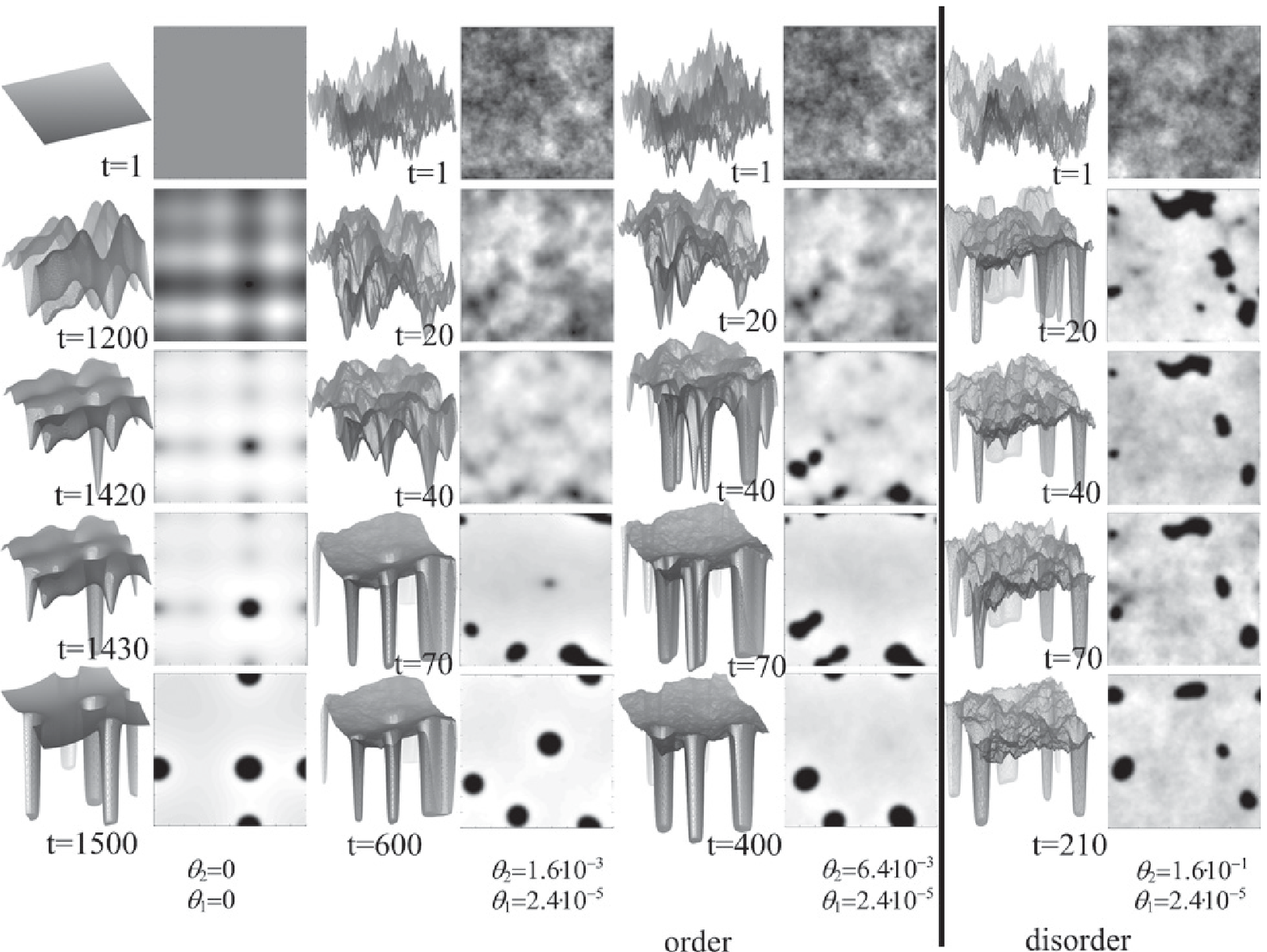}\\
\caption{ The evolution of spatial Turing patterns at different noise
intensities away from the deterministic transition point. See explanation
for Fig. 7.}
\end{figure}

Fig. 8 demonstrates the evolution of spatial patterns arising spontaneously
away from the Turing bifurcation point. The initial conditions correspond to
the homogeneous stationary state of the system without initial perturbation.
The boundary conditions are periodic. The domain of integration is square
200$\times $200. The modeling parameters are $m_{0} / r_{0} = 0.490$, $a_{0} / r_{0} =  8$, $g^{2} / r_{0} = {\rm 2}.{\rm 0}{\rm 5}{\rm 6}$, $f = {\rm 0}.093$, $h = {\rm 0}.857$, $b = 11.905$,
$D_{2} / D_{1} = 1000$.

It is obviously from Fig. 8, that all the regularities described above are
retained in the evolution of the system away from the bifurcation point. It
must be emphasized here, that the "order-disorder" transition in this case
takes place at greater values of noise intensity than in the vicinity of the
bifurcation point.

Fig. 9 gives a clear idea of the system's behavior at high level of noise.
As opposed to pure noise-induced transition here the system's random
behavior is conditioned by fast formation and destruction of a great number
of patterns having random contours and arising in random places.

\begin{figure}
\centering
\includegraphics[width=6.32in]{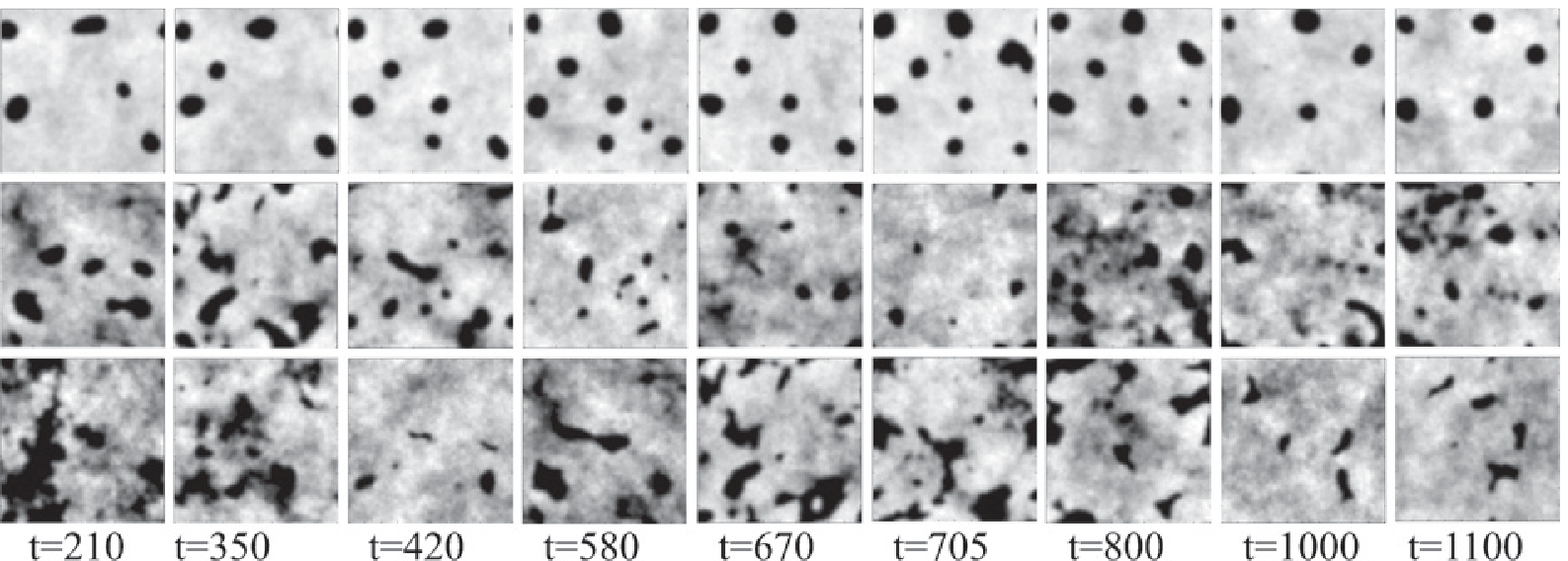}\\
\caption{The evolution of spatial Turing patterns in a strong noise. The
figure shows a strongly irregular behavior (disorder) of the system (\ref{eq22}).
The top horizontal row: \textit{$\theta $}$_{{\rm 2}}$=0.16. The center horizontal row:
\textit{$\theta $}$_{{\rm 2}}$=0.64. The bottom horizontal row: \textit{$\theta $}$_{{\rm 2}}$=1.44. \textit{$\theta $}$_{{\rm 1}}$=2.40$\cdot $10$^{{\rm -} {\rm 5}}$, $r_{f{\rm 1}}=r_{f{\rm 2}}$=1.}
\end{figure}

In the light of the theory developed in our study such behavior is due to
the fact, that the region of the system's instability considerable expands at
high level of noise and pattern formation takes place owing to cooperation
and competition of a very large number of unstable modes with different
conditions of resonant interaction.

We have also carried out simulation of system (22) evolution in the
subcritical region. The following parameters $m_{0} / r_{0} = \mu _{1} =
0.510$, $a_{0} / r_{0} = \mu _{2} = 8$, $g^{2} / r_{0} = {\rm 2}.{\rm
0}{\rm 6}$, $f = {\rm 0}.093$, $h = {\rm 0}.857$, $b = 11.905$, $D_{2} /
D_{1} = 1000$, $\theta _{{\rm 1}}$=2.6$\cdot $10$^{{\rm -} {\rm 3}}, \theta
_{{\rm 2}}$=2.56$\cdot $10$^{{\rm -} {\rm 2}}$; $k_{{\rm f}{\rm 1}} = k_{{\rm
f}{\rm 2}} = 0.5$, the domain 200$\times $200, and periodic boundary
conditions are chosen for the modeling.

Fig. 10 presents the evolution of $x_{{\rm 1}}$ state function distribution
for the instants of time t = 1, 927, 940, 950 and 1120.

\begin{figure}
\centering
\includegraphics[width=5.98in]{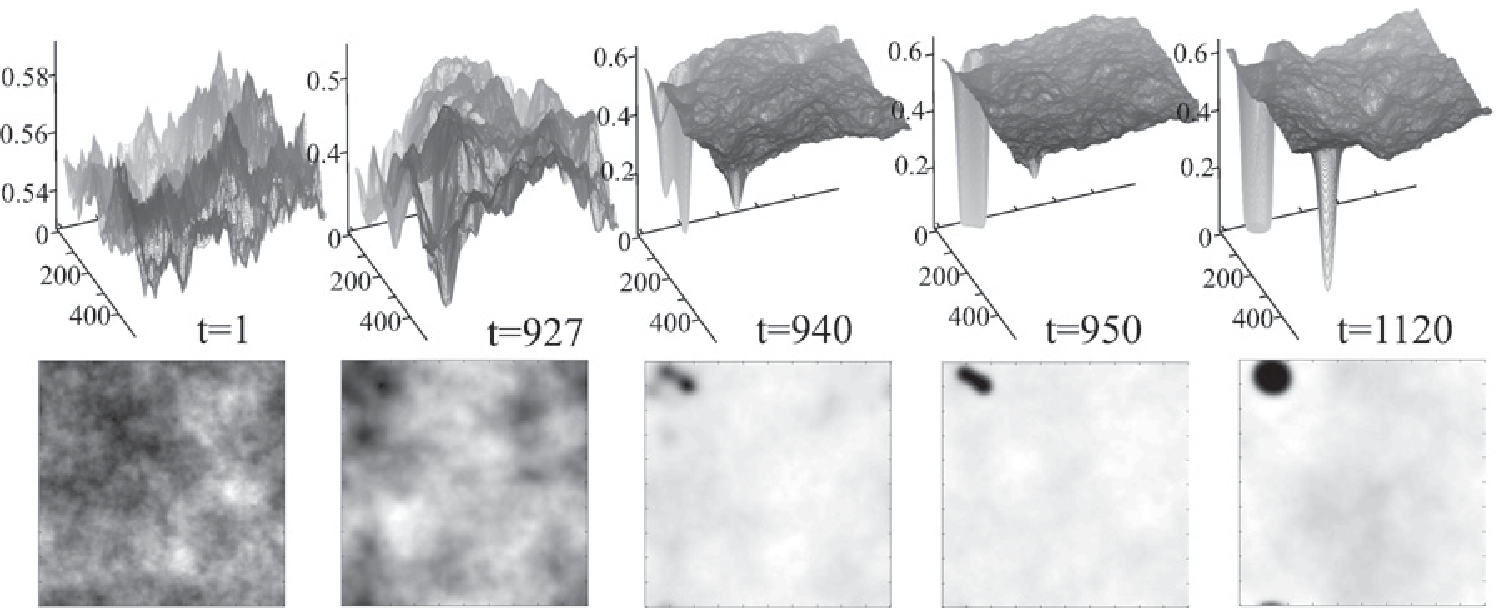}\\
\caption{ Noise-induced parametric excitation of spatial Turing pattern in
the subcritical region.}
\end{figure}

The pattern arising under parametric instability has a ``solitonlike'' shape
(see Fig. 10 t=1120). From Fig. 10 it is also clear, that the duration of the
process of destroying the homogeneous state in the subcritical region (see
Fig. 10 t=1-927) is considerably longer than that in the supercritical
region (see Fig. 8 t=1-40 in the case of order). As predicted in section
V.A, the initiation of inhomogeneous state in the subcritical region may be
seen only as a result of prolonged observation of the system's evolution.

Thus, the results of simulation of system (\ref{eq22}) evolution qualitatively fully
confirm the theoretical conclusions of section V. A comparison of Fig. 5
and Fig. 7 makes it possible to establish a satisfactory quantitative
correspondence between the theory and the numerical experiment in the
vicinity of the Turing bifurcation point.

\section{Conclusion}

In our study a theory is developed, that makes it possible to predict and
analyze in detail different noise-induced effects occurring in open nonlinear
distributed multicomponent multi-dimensional systems both in the vicinity of
and away from the deterministic transition point. The prediction and
analysis are based on the unified point of view of the concept of order
parameters.

Stochastic equations have been obtained for amplitudes of unstable modes
(order parameters), as well as dispersion equation for averaged amplitudes
of unstable modes. The dependence of eigenvalues of unstable mode averaged
amplitudes on wavenumbers, noise intensity, and correlation length has
been found analytically. The Fokker-Planck equation for the order parameters
of the systems under study has been obtained. Its stationary solution for a
critical mode has been obtained in an explicit form.

As our theory predicts, the increments of unstable modes change, the
system's instability region is extended, the conditions of mode resonance
interaction change, and noise-induced parametric instability occurs in
external noise. The destruction of a homogeneous state and the pattern
formation take place faster than in the deterministic case. Our theory
points to the existence of a noise-induced "disorder - order - disorder"
phase transition in systems of the type discussed.

The advantages of the approach developed lies in the fact, that it applies to
multicomponent multidimensional systems. The stationary solution to the
Fokker-Planck equation for the critical order parameter is written in an
explicit form. The approach applies to a wider class of functions $P_{1j}
(x_{1} ,x_{2} ,\chi _{m + 1} ,...,\chi _{{n}'} )$, $P_{2j} (x_{1} ,x_{2}
,\eta _{l + 1} ,...,\eta _{{s}'} )$ including those with a discontinuity of
the second kind. Moreover, the approach suggested does not contain an
arbitrary element connected with the discretization of system's continuous
space. It can also be used in the case of noise with finite characteristic
spatial and temporal scales.

The numerical analysis of a specific system of the type considered
qualitatively confirms the theoretical conclusions and provides a satisfactory
quantitative correspondence between the theory and the numerical experiment in
the vicinity of the bifurcation point.

\appendix

\section{}

Functions $\sigma _{{j}'{j}''}^{(j)} ({\rm {\bf k}},{\rm {\bf {k}'}},{\rm
{\bf {k}''}})$, $\sigma _{{j}'{j}''{j}'''}^{(j)} ({\rm {\bf k}},{\rm {\bf
{k}'}},{\rm {\bf {k}''}},{\rm {\bf {k}'''}})$, $\varepsilon _{\varphi
,{j}'}^{(j)} ({\rm {\bf k}},{\rm {\bf {k}'}})$, $\varepsilon _{\varphi
,{j}'{j}''}^{(j)} ({\rm {\bf k}},{\rm {\bf {k}'}},{\rm {\bf {k}''}})$
introduced in equations (\ref{eq10}):

\[
\sigma _{{j}'{j}''}^{(j)} ({\rm {\bf k}},{\rm {\bf {k}'}},{\rm {\bf {k}''}})
= {\sum\limits_{\varepsilon ,\mu ,\nu}  {g_{\varepsilon ,\mu \nu} ^{(2)}
O_{\varepsilon} ^{ * (j)} ({\rm {\bf k}})O_{\mu} ^{({j}')} ({\rm {\bf
{k}'}})O_{\nu} ^{({j}'')} ({\rm {\bf {k}''}})}} ,
\]

\[
\sigma _{{j}'{j}''{j}'''}^{(j)} ({\rm {\bf k}},{\rm {\bf {k}'}},{\rm {\bf
{k}''}},{\rm {\bf {k}'''}}) = {\sum\limits_{\varepsilon ,\mu ,\nu ,\kappa}
{g_{\varepsilon ,\mu \nu \kappa} ^{(3)} O_{\varepsilon} ^{ * (j)} ({\rm {\bf
k}})O_{\mu} ^{({j}')} ({\rm {\bf {k}'}})O_{\nu} ^{({j}'')} ({\rm {\bf
{k}''}})O_{\kappa} ^{({j}''')} ({\rm {\bf {k}'''}})}} ,
\]

\[
\varepsilon _{\varphi ,{j}'}^{(j)} ({\rm {\bf k}},{\rm {\bf {k}'}}) =
{\sum\limits_{\mu}  {p_{\varphi ,\mu} ^{(1)} O_{\varphi} ^{ * (j)} ({\rm
{\bf k}})O_{\mu} ^{({j}')} ({\rm {\bf {k}'}})}} ,
\]

\[
\varepsilon _{\varphi ,{j}'{j}''}^{(j)} ({\rm {\bf k}},{\rm {\bf {k}'}},{\rm
{\bf {k}''}}) = {\sum\limits_{\mu ,\nu}  {p_{\varphi ,\mu \nu} ^{(2)}
O_{\varphi} ^{ * (j)} ({\rm {\bf k}})O_{\mu} ^{({j}')} ({\rm {\bf
{k}'}})O_{\nu} ^{({j}'')} ({\rm {\bf {k}''}})}} .
\]

\section{}

Function $\omega ({\rm {\bf k}}_{u} ,{\rm {\bf {k}'}}_{u} ,{\rm {\bf
{k}''}}_{u} ,{\rm {\bf {k}'''}}_{u} ,{\rm {\bf k}}_{u} - {\rm {\bf
{k}'}}_{u} )$ and others introduced in equations (\ref{eq14}):

\[
\omega ({\rm {\bf k}}_{u} ,{\rm {\bf {k}'}}_{u} ,{\rm {\bf {k}''}}_{u} ,{\rm
{\bf {k}'''}}_{u} ,{\rm {\bf k}}_{u} - {\rm {\bf {k}'}}_{u} ) = \sigma
_{111}^{(1)} ({\rm {\bf k}}_{u} ,{\rm {\bf {k}'}}_{u} ,{\rm {\bf {k}''}}_{u}
,{\rm {\bf {k}'''}}_{u} ) -
\]
\[
 - {\frac{{[\sigma _{11}^{(1)} ({\rm {\bf k}}_{u} ,{\rm {\bf {k}'}}_{u}
,{\rm {\bf k}}_{s} ) + \sigma _{11}^{(1)} ({\rm {\bf k}}_{u} ,{\rm {\bf
k}}_{s} ,{\rm {\bf {k}'}}_{u} )]}}{{\lambda _{1} ({\rm {\bf k}}_{s}
)}}}\sigma _{11}^{(1)} ({\rm {\bf k}}_{s} ,{\rm {\bf {k}''}}_{u} ,{\rm {\bf
{k}'''}}_{u} ) -
\]
\[
 - {\frac{{[\sigma _{12}^{(1)} ({\rm {\bf k}}_{u} ,{\rm {\bf {k}'}}_{u}
,{\rm {\bf k}}_{s} ) + \sigma _{21}^{(1)} ({\rm {\bf k}}_{u} ,{\rm {\bf
k}}_{s} ,{\rm {\bf {k}'}}_{u} )]}}{{\lambda _{2} ({\rm {\bf k}}_{s}
)}}}\sigma _{11}^{(2)} ({\rm {\bf k}}_{s} ,{\rm {\bf {k}''}}_{u} ,{\rm {\bf
{k}'''}}_{u} ),
\]

\[
\zeta _{\varphi ,\psi ,{\varphi} '} ({\rm {\bf k}}_{u} ,{\rm {\bf k}}_{s} )
= \varepsilon _{\varphi ,\psi} ^{(1)} ({\rm {\bf k}}_{u} ,{\rm {\bf k}}_{s}
){\frac{{O_{{\varphi} '}^{ * (\psi )} ({\rm {\bf k}}_{s} )}}{{\lambda _{\psi
} ({\rm {\bf k}}_{s} )}}}p_{{\varphi} '}^{(0)} ,
\]

\[
\eta _{\varphi}  ({\rm {\bf k}}_{u} ,{\rm {\bf {k}'}}_{u} ) = \varepsilon
_{\varphi ,1}^{(1)} ({\rm {\bf k}}_{u} ,{\rm {\bf {k}'}}_{u} ) -
\]
\[
-{\sum\limits_{\psi = 1}^{2} {[\sigma _{1\psi} ^{(1)} ({\rm {\bf k}}_{u}
,{\rm {\bf {k}'}}_{u} ,{\rm {\bf k}}_{u} - {\rm {\bf {k}'}}_{u} ) + \sigma
_{\psi 1}^{(1)} ({\rm {\bf k}}_{u} ,{\rm {\bf k}}_{u} - {\rm {\bf {k}'}}_{u}
,{\rm {\bf {k}'}}_{u} )]{\frac{{O_{\varphi} ^{ * (\psi )} ({\rm {\bf k}}_{u}
- {\rm {\bf {k}'}}_{u} )}}{{\lambda _{\psi}  ({\rm {\bf k}}_{u} - {\rm {\bf
{k}'}}_{u} )}}}p_{\varphi} ^{(0)}}}  ,
\]

\[
\nu _{\varphi}  ({\rm {\bf k}}_{u} ,{\rm {\bf {k}'}}_{u} ,{\rm {\bf
{k}''}}_{u} ) = \varepsilon _{\varphi ,11}^{(1)} ({\rm {\bf k}}_{u} ,{\rm
{\bf {k}'}}_{u} ,{\rm {\bf {k}''}}_{u} ) -
\]
\[
-{\sum\limits_{\psi = 1}^{2} {[\sigma _{1\psi} ^{(1)} ({\rm {\bf k}}_{u}
,{\rm {\bf {k}'}}_{u} ,{\rm {\bf k}}_{u} - {\rm {\bf {k}'}}_{u} ) + \sigma
_{\psi 1}^{(1)} ({\rm {\bf k}}_{u} ,{\rm {\bf k}}_{u} - {\rm {\bf {k}'}}_{u}
,{\rm {\bf {k}'}}_{u} )]{\frac{{\varepsilon _{\varphi ,1}^{(\psi )} ({\rm
{\bf k}}_{u} - {\rm {\bf {k}'}}_{u} ,{\rm {\bf {k}''}}_{u} )}}{{\lambda
_{\psi}  ({\rm {\bf k}}_{u} - {\rm {\bf {k}'}}_{u} )}}}}}  -
\]
\[
 - {\sum\limits_{\psi ,\varphi = 1}^{2} {\varepsilon _{\varphi ,\psi} ^{(1)}
({\rm {\bf k}}_{u} ,{\rm {\bf {k}'}}_{u} + {\rm {\bf {k}''}}_{u}
){\frac{{\sigma _{11}^{(\psi )} ({\rm {\bf k}}_{u} ,{\rm {\bf {k}'}}_{u}
,{\rm {\bf {k}''}}_{u} )}}{{\lambda _{\psi}  ({\rm {\bf {k}'}}_{u} + {\rm
{\bf {k}''}}_{u} )}}}}} ,
\]

\[
A_{\varphi ,\psi ,{\varphi} '} ({\rm {\bf k}}_{u} ,{\rm {\bf k}}_{s} ,{\rm
{\bf {k}'}}_{u} ) = \varepsilon _{\varphi ,\psi} ^{(1)} ({\rm {\bf k}}_{u}
,{\rm {\bf k}}_{s} ){\frac{{\varepsilon _{{\varphi} ',1}^{(\psi )} ({\rm
{\bf k}}_{s} ,{\rm {\bf {k}'}}_{u} )}}{{\lambda _{\psi}  ({\rm {\bf k}}_{s}
)}}},
\]

\[
B_{\varphi ,\psi ,{\varphi} '} ({\rm {\bf k}}_{u} ,{\rm {\bf k}}_{s} ,{\rm
{\bf {k}'}}_{u} ) = [\varepsilon _{\varphi ,1\psi} ^{(1)} ({\rm {\bf k}}_{u}
,{\rm {\bf {k}'}}_{u} ,{\rm {\bf k}}_{s} ) + \varepsilon _{\varphi ,\psi
1}^{(1)} ({\rm {\bf k}}_{u} ,{\rm {\bf k}}_{s} ,{\rm {\bf {k}'}}_{u}
)]{\frac{{O_{{\varphi} '}^{ * (\psi )} ({\rm {\bf k}}_{s} )}}{{\lambda
_{\psi}  ({\rm {\bf k}}_{s} )}}}p_{{\varphi} '}^{(0)} ,
\]

\[
C_{\varphi ,\psi ,{\varphi} '} ({\rm {\bf k}}_{u} ,{\rm {\bf k}}_{s} ,{\rm
{\bf {k}'}}_{u} ,{\rm {\bf {k}''}}_{u} ) = \varepsilon _{\varphi ,\psi
}^{(1)} ({\rm {\bf k}}_{u} ,{\rm {\bf k}}_{s} ){\frac{{\varepsilon
_{{\varphi} ',11}^{(\psi )} ({\rm {\bf k}}_{s} ,{\rm {\bf {k}'}}_{u} ,{\rm
{\bf {k}''}}_{u} )}}{{\lambda _{\psi}  ({\rm {\bf k}}_{s} )}}},
\]

\[
D_{\varphi ,\psi ,{\varphi} '} ({\rm {\bf k}}_{u} ,{\rm {\bf {k}'}}_{u}
,{\rm {\bf k}}_{s} ,{\rm {\bf {k}''}}_{u} ) = [\varepsilon _{\varphi ,1\psi
}^{(1)} ({\rm {\bf k}}_{u} ,{\rm {\bf {k}'}}_{u} ,{\rm {\bf k}}_{s} ) +
\varepsilon _{\varphi ,\psi 1}^{(1)} ({\rm {\bf k}}_{u} ,{\rm {\bf k}}_{s}
,{\rm {\bf {k}'}}_{u} )]{\frac{{\varepsilon _{{\varphi} ',1}^{(\psi )} ({\rm
{\bf k}}_{s} ,{\rm {\bf {k}''}}_{u} )}}{{\lambda _{\psi}  ({\rm {\bf k}}_{s}
)}}},
\]

\[
E_{\psi ,\varphi}  ({\rm {\bf k}}_{u} ,{\rm {\bf {k}'}}_{u} ,{\rm {\bf
k}}_{u} - {\rm {\bf {k}'}}_{u} ,{\rm {\bf {k}''}}_{u} ,{\rm {\bf
{k}'''}}_{u} ) =
\]
\[
=[\sigma _{1\psi} ^{(1)} ({\rm {\bf k}}_{u} ,{\rm {\bf {k}'}}_{u} ,{\rm {\bf
k}}_{u} - {\rm {\bf {k}'}}_{u} ) + \sigma _{\psi 1}^{(1)} ({\rm {\bf k}}_{u}
,{\rm {\bf k}}_{u} - {\rm {\bf {k}'}}_{u} ,{\rm {\bf {k}'}}_{u}
)]{\frac{{\varepsilon _{\varphi ,11}^{(\psi )} ({\rm {\bf k}}_{u} - {\rm
{\bf {k}'}}_{u} ,{\rm {\bf {k}''}}_{u} ,{\rm {\bf {k}'''}}_{u} )}}{{\lambda
_{\psi}  ({\rm {\bf k}}_{u} - {\rm {\bf {k}'}}_{u} )}}},
\]

\[
F_{\psi ,\varphi}  ({\rm {\bf k}}_{u} ,{\rm {\bf {k}'}}_{u} ,{\rm {\bf
{k}''}}_{u} ,{\rm {\bf {k}'''}}_{u} ) =
\]
\[
=[\varepsilon _{\varphi ,1\psi} ^{(1)} ({\rm {\bf k}}_{u} ,{\rm {\bf
{k}'}}_{u} ,{\rm {\bf {k}''}}_{u} + {\rm {\bf {k}'''}}_{u} ) + \varepsilon
_{\varphi ,\psi 1}^{(1)} ({\rm {\bf k}}_{u} ,{\rm {\bf {k}''}}_{u} + {\rm
{\bf {k}'''}}_{u} ,{\rm {\bf {k}'}}_{u} )]{\frac{{\sigma _{11}^{(\psi )}
({\rm {\bf {k}''}}_{u} + {\rm {\bf {k}'''}}_{u} ,{\rm {\bf {k}''}}_{u} ,{\rm
{\bf {k}'''}}_{u} )}}{{\lambda _{\psi}  ({\rm {\bf {k}''}}_{u} + {\rm {\bf
{k}'''}}_{u} )}}},
\]

\[
G_{\psi ,\varphi ,{\varphi} '} ({\rm {\bf k}}_{u} ,{\rm {\bf {k}'}}_{u}
,{\rm {\bf k}}_{s} ,{\rm {\bf {k}''}}_{u} ,{\rm {\bf {k}'''}}_{u} ) =
\]
\[
=[\varepsilon _{\varphi ,1\psi} ^{(1)} ({\rm {\bf k}}_{u} ,{\rm {\bf
{k}'}}_{u} ,{\rm {\bf k}}_{s} ) + \varepsilon _{\varphi ,\psi 1}^{(1)} ({\rm
{\bf k}}_{u} ,{\rm {\bf k}}_{s} ,{\rm {\bf {k}'}}_{u} )]{\frac{{\varepsilon
_{{\varphi} ',11}^{(\psi )} ({\rm {\bf k}}_{s} ,{\rm {\bf {k}''}}_{u} ,{\rm
{\bf {k}'''}}_{u} )}}{{\lambda _{\psi}  ({\rm {\bf k}}_{s} )}}}.
\]

\end{document}